\documentclass[11pt,a4paper]{article}
\pdfoutput=1 
\usepackage{jheppub}
\usepackage{graphicx}
\usepackage{amssymb,amsmath,amsbsy}
\usepackage{psfrag}
\usepackage{tikzfeynman}

\newcommand{\fref}[1]{figure~\ref{fig:#1}} 
\newcommand{\eref}[1]{eq.~\eqref{eq:#1}}

\newcommand{\aref}[1]{Appendix~\ref{app:#1}}
\newcommand{\sref}[1]{section~\ref{sec:#1}}
\newcommand{\cref}[1]{Chapter~\ref{ch:.#1}}

\newcommand{\nnl}{\nonumber \\}

\newcommand{\beq}{\begin{equation}} 
\newcommand{\eeq}{\end{equation}} 
\newcommand{\bea}{\begin{eqnarray}}  
\newcommand{\eea}{\end{eqnarray} }   
\newcommand{\bal}{\begin{align}}
\newcommand{\eal}{\end{align}}   
\newcommand{\bi}{\begin{itemize}}  
\newcommand{\ei}{\end{itemize}}  
\newcommand{\ben}{\begin{enumerate}}  
\newcommand{\een}{\end{enumerate}}  
\newcommand{\bc}{\begin{center}}
\newcommand{\ec}{\end{center}} 
\newcommand{\bt}{\begin{table}}
\newcommand{\et}{\end{table}}  
\newcommand{\btb}{\begin{tabular}}
\newcommand{\etb}{\end{tabular}}

\newcommand{\cO}{{\mathcal O}} 
 
\newcommand{\cL}{{\mathcal L}} 
\newcommand{\cl}{{\mathcal L}} 
\newcommand{\cM}{{\mathcal M}}

\newcommand{\re}{{\mathrm{Re}} \,}
\newcommand{\im}{{\mathrm{Im}} \,}

\def\hc{{\rm h.c.}} 

\newcommand{\eps}{\epsilon}

\newcommand{\be}{\begin{equation}}
\newcommand{\ee}{\end{equation}}
\newcommand{\ba}{\begin{eqnarray}}
\newcommand{\ea}{\end{eqnarray}}

\author[a]{Herm\`es B\'elusca-Ma\"ito,}
\author[b]{Adam Falkowski,}
\author[c]{Duarte Fontes,}
\author[c]{Jorge.~C.~Rom\~{a}o,}
\author[c]{Jo\~{a}o~P.~Silva}
\affiliation[a]{University of Zagreb, Department of Physics, Bijeni\v{c}ka cesta 32, 10000
Zagreb, Croatia}
\affiliation[b]{Laboratoire de Physique Th\'{e}orique, CNRS, Univ. Paris-Sud, Universit\'{e}  Paris-Saclay, 91405 Orsay, France}
\affiliation[c]{Departamento de F\'{\i}sica e
Centro de F\'{\i}sica Te\'{o}rica de Part\'{\i}culas (CFTP), Instituto Superior T\'{e}cnico, Universidade de Lisboa, 1049-001 Lisboa, Portugal}
\emailAdd{hbelusca@phy.hr}
\emailAdd{adam.falkowski@th.u-psud.fr}
\emailAdd{duartefontes@tecnico.ulisboa.pt}
\emailAdd{jorge.romao@tecnico.ulisboa.pt}
\emailAdd{jpsilva@cftp.tecnico.ulisboa.pt}

\title{CP violation in 2HDM and EFT: the $ZZZ$ vertex}

\abstract{
We study the CP violating $ZZZ$ vertex in the two-Higgs doublet model,  which is a probe of a Jarlskog-type invariant in the extended Higgs sector. 
The form factor $f_4^Z$ is evaluated at one loop in a general $R_\xi$ gauge and its magnitude is estimated in the realistic parameter space.
Then we turn to the decoupling limit of the two-Higgs doublet model, where the extra scalars are heavy and the physics can be described by the Standard Model supplemented by higher-dimensional operators.
The leading operator contributing to $f_4^Z$ at one loop is identified. 
The CP violating $ZZZ$ vertex is not generated in the effective theory by dimension-8 operators,  but  instead arises only at the dimension-12 level, which implies an additional suppression by powers of the  heavy Higgs mass scale.    
}

\arxivnumber{}

\begin{document}
\begin{flushright}
CFTP/17-007 \\ 
LPT-Orsay-17-45 \\ 
ZTF-EP-17-10
\end{flushright}
\maketitle

\section{\label{sec:intro}Introduction}

The LHC experiments announced in 2012 the discovery of a neutral scalar (h) of  mass 125 GeV 
\cite{Aad:2012tfa,Chatrchyan:2012xdj}, consistent with the 1964 prediction of a Higgs Boson
as a by-product of the spontaneous symmetry breaking of a gauge symmetry
\cite{Higgs:1964ia, Higgs:1964pj, Englert:1964et, Guralnik:1964eu}.
This opened a very exciting program addressing two fundamental questions:
i) how many scalars are there?,
and ii) do the couplings of the 125 GeV state conform to
the prediction in the Standard Model (SM) of electroweak interactions
\cite{Glashow:1961tr, Weinberg:1967tq}?  
Thus far, there is no definite sign of an inconsistency with the SM. 

A model independent way to interpret the data is provided
by the SM effective field theory (SMEFT), where one allows for all operators constructed from the
SM fields, organized in an expansion
\be
{\cal L}_\textrm{eff}
=
{\cal L}_\textrm{SM}
+
\sum_{D=5}^\infty
\left(
\sum_i
\frac{c_i^{(D)}}{\Lambda^{D-4}}
{\cal O}_i^{(D)}
\right).
\ee
${\cal L}_\textrm{SM}$ is the SM Lagrangian,
$\Lambda$ is the mass scale at which new
degrees of freedom become propagating,
each ${\cal O}_i^{(D)}$ is an
$SU(3) \times SU(2) \times U(1)$ invariant operator
of dimension $D$, and $c_i^{(D)}$ the corresponding
Wilson coefficient.
The new operators modify the interaction strength of the SM particles, or introduce new interactions that are not predicted within the SM.
Identifying their presence in the interaction Lagrangian would not only be an evidence of new physics, 
but would also give indirect hints about the mass scale and degrees of freedom of the underlying theory beyond the SM.

A complementary approach consists in investigating the constraints that the data place  on well motivated theories. 
One simple example consists in adding one more scalar doublet to the SM.
Two Higgs doublet models (2HDMs) are interesting because they
contain many properties which one may find in more complicated theories,
such as the presence of extra neutral scalars, charged scalars,
CP-odd or admixtures of CP-even and CP-odd scalars,
the possibilities for spontaneous CP violation or
flavour changing neutral scalar interactions,
among others \cite{Gunion:1989we,Branco:2011iw}.
Here we focus on the so-called ``complex 2HDM''
(C2HDM)~\cite{Ginzburg:2002wt, Khater:2003wq, ElKaffas:2007rq, ElKaffas:2006gdt, Grzadkowski:2009iz, Arhrib:2010ju, Barroso:2012wz, Fontes:2014xva}.

2HDMs can be approximated by the SMEFT at energies below the mass scale of the new scalars, 
where the indirect effects of the new scalars are represented by a tower of the higher-dimensional operators in the Lagrangian.   
There has been some interest in the matching between the 2HDM parameters and
the SMEFT Wilson coefficients~\cite{Gorbahn:2015gxa, Brehmer:2015rna, Egana-Ugrinovic:2015vgy,
Freitas:2016iwx, Belusca-Maito:2016dqe,Corbett:2017ieo,Karmakar:2017yek}.
This exercise allows one to get some intuition about the pattern of operators expected from realistic extensions of the SM, and to identify the leading new physics effects in a model-independent language.  
In this paper we discuss how CP violation of the C2HDM is manifested in the SMEFT.    
More precisely, we concentrate on the CP violating $ZZZ$ vertex, which
appears at one loop in the C2HDM.   
This is an especially interesting observable because it measures
directly a Jarlskog-type invariant in the Higgs-gauge sector
\cite{Grzadkowski:2016lpv}, first introduced in \cite{Lavoura:1994fv,Botella:1994cs}.
At the technical level, the C2HDM computation involves loops with both heavy and light
particles, which require special care when matching to the low-energy effective theory \cite{delAguila:2016zcb,Henning:2016lyp,Ellis:2016enq,Fuentes-Martin:2016uol,Zhang:2016pja,Ellis:2017jns}. 
In fact, we will show that the effective description of the CP violating $ZZZ$ vertex is quite non-trivial in this case.
Generally, the lowest order in the SMEFT expansion where the  $ZZZ$ vertex may appear is dimension-8 $(\cO(\Lambda^{-4}))$ \cite{Gounaris:2006zm,Degrande:2013kka}. 
However, in turns out that in the SMEFT matched to the C2HDM at one loop the CP-violating $ZZZ$ vertex is only generated at $(\cO(\Lambda^{-8}))$, that is by a dimension-12 operator.  

Our paper is organized as follows.
In~\sref{ZZZ} we review the observable form factors  associated  with the $ZZZ$ vertex. 
Our  C2HDM notation and conventions  are summarized  in~\sref{notation}. 
Section~\ref{sec:ZZZ_C2HDM} contains the calculation of the CP-violating contribution to the $ZZZ$ vertex in a general $R_\xi$ gauge, which we compare with that in the previous literature~\cite{Grzadkowski:2016lpv}.
The corresponding calculation and the CP violating operator in the SMEFT is discussed in section~\ref{sec:eft2hdm}.
Finally, \sref{conclusions} presents our conclusions. 
{  Some technical details concerning  approximation of the loop integrals using the method of regions are given in \aref{mr}, while derivation of the CP-violating dimension-12  SMEFT operators using gauge invariant functional methods is given in \aref{eft}.}

\section{\label{sec:ZZZ} The $ZZZ$ vertex}

\begin{figure}
\bc
\begin{tikzpicture}[line width=1.5 pt, scale=1.3]
	\draw[vector] (180:2)--(0,0); 
	\draw[vector] (60:2)--(0,0);
	\draw[vector] (300:2)--(0,0);
	\node at (180:2.2) {$\bf Z_\mu$};
	\node at (170:1.8) {$q$};
         \draw[->] (-1.6,0.3) -- (-1.4,0.3);
	\node at (60:2.2) {$\bf Z_\alpha$};
	\node at (42:1.5) {$p_1$};
	\draw[->] (0.8,1.0) -- (0.9,1.2);
	\node at (300:2.2) {$\bf Z_\beta$};	
	\node at (318:1.5) {$p_2$};
         \draw[->] (0.8,-1.0) -- (0.9,-1.2);
	\draw[fill=black] (0,0) circle (.12cm);
          	
\end{tikzpicture}
 \ec 
 \caption{Conventions for the $ZZZ$ vertex $\Gamma_{\mu \alpha \beta}$.
 \label{fig:zzz}}
 \end{figure}
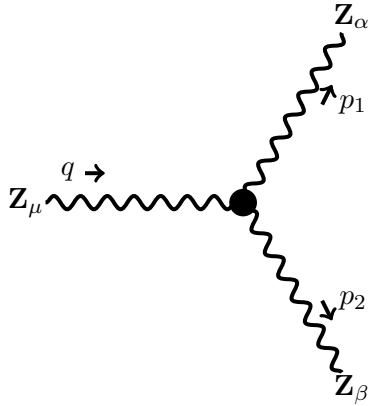

We start by reviewing the formalism to describe the effective $Z^3$ vertex \cite{Hagiwara:1986vm,Gounaris:1999kf}. 
Consider the diagram in \fref{zzz}  with two on-shell $Z$ bosons characterized by outgoing  
4-momenta $p_1$, $p_2$ and polarization vectors $\eps(p_1)$, $\eps(p_2)$,
and an {\em off-shell} $Z$ boson with the incoming momentum $q = p_1+ p_2$.    
It follows that $p_1^2 = p_2^2 = m_Z^2$,   $p_1 p_2 = q^2/2 - m_Z^2$, 
$q p_1 = q p_2 = q^2/2$.  
The blob in \fref{zzz} may represent a contact interaction or particles running in loops.  
The Lorentz and Bose symmetries constrain the $Z^3$ vertex function $\Gamma_{\mu \alpha \beta}$ to the following form: 
\bea
\label{eq:myZZZvertex}
\hspace{-1cm}
i \Gamma_{\mu \alpha \beta} & =&  
- {q^2 - m_Z^2 \over m_Z^2}  e f_4^Z(q^2) 
\left (\eta_{\mu \alpha} p_{1,\beta}  + \eta_{\mu \beta} p_{2,\alpha} \right )  
- {q^2 - m_Z^2 \over m_Z^2}  e f_5^Z(q^2)  \epsilon_{\mu \alpha \beta \rho} (p_1 - p_2)^{\rho} 
\nnl & + & 
 \tilde f_1(q^2)  (\eta_{\mu \alpha} p_{2,\beta}  + \eta_{\mu \beta} p_{1,\alpha} ) 
+  \tilde  f_2(q^2)  \eta_{\alpha \beta} q_{\mu}  
\nnl & + &   \tilde f_3(q^2)  q_{\mu}  p_{1,\beta} p_{2,\alpha}   
+   \tilde f_4(q^2)  q_{\mu}  p_{1,\alpha} p_{2,\beta} 
+   \tilde f_5(q^2)  q_{\mu}  (p_{1,\alpha} p_{1,\beta}   + p_{2,\alpha} p_{2,\beta}  ) . 
\eea    
In the first line we have pulled out a function of $q^2$ from the form factors so as to match the standard notation of Ref.~\cite{Gounaris:1999kf}.\footnote{
For s-channel production of an on-shell $ZZ$ pair from a conserved current our vertex parametrization in  \eref{myZZZvertex} and the one in Ref.~\cite{Gounaris:1999kf} both lead to the same amplitude \eref{mZZ}. 
The two parametrizations differ only at the level of non-physical unmeasurable form factors.   
}  

The form factor $f_4^Z(q^2)$ corresponds to a $C$-odd and $P$-even  (thus $CP$-odd) interaction. 
One way to see this  is to  note that an effective Lagrangian with the $Z^3$ interaction,  
\beq
\label{eq:lZZZ}
\cl_{\rm eff}  \supset  {\tilde \kappa_{ZZZ} \over m_Z^2} 
\partial_\mu Z_\nu \partial^\mu Z^\rho  \partial_\rho Z^\nu,  
\eeq 
leads to the tree-level vertex in \eref{myZZZvertex} with  $f_4^Z(q^2) =  \tilde \kappa_{ZZZ}$.     
The CP properties are then easily obtained given $C$ acting as $Z_\mu \to - Z_\mu$ and $P$ acting as 
$Z_0 \to Z_0$, $Z_i \to - Z_i$, $\partial_0 \to \partial_0$, $\partial_i \to - \partial_i$. 
By similar argument one shows that  $f_5^Z(q^2)$ corresponds to a $C$-odd and $P$-odd  (thus $CP$-even) interaction.

It is important to stress that the full vertex $\Gamma_{\mu \alpha \beta}$ is not an observable. 
Nevertheless, the form factors $f_4^Z$ and $f_5^Z$ can be related to observable quantities in the following sense. 
Consider the amplitude to produce a pair of $Z$ bosons. 
This process will receive a contribution from the diagram with an intermediate off-shell $Z$ boson in the s-channel:   $\cM_{f \bar f \to ZZ} =  \cM_{f \bar f \to ZZ}^{(s)} + \dots$, where the dots stand for other contributions. 
The s-channel can be written as  
$\cM_{f \bar f \to ZZ}^{(s)} =  {1 \over q^2 - m_Z^2} \Gamma_{\mu \alpha \beta} \eps^\alpha(p_1) \eps^\beta(p_2)  j^\mu(q)$,  where $j^\mu$ is the $f$ current to which the $Z$ boson couples in the Lagrangian. 
We assume that $j^\mu$ is conserved, $q_\mu j^\mu = 0$,  which is the case in the relevant situation of $q \bar q$ and $e^+ e^-$ collisions in the limit where the fermions are treated as massless.
Moreover, $p_1^\alpha \eps_\alpha(p_1) = p_2^\beta \eps_\beta(p_2) =0$.  
Then the s-channel part of the $ZZ$ production amplitude reduces to 
\beq
\label{eq:mZZ}
\cM_{f \bar f \to ZZ}^{(s)}  =  -{1 \over m_Z^2} \left [ f_4^Z(q^2) \left (\eta_{\mu \alpha} p_{1,\beta}  + \eta_{\mu \beta} p_{2,\alpha} \right )
+ f_ 5^Z(q^2)\epsilon_{\mu \alpha \beta \rho} (p_1 - p_2)^{\rho} 
 \right ]  \eps^\alpha(p_1) \eps^\beta(p_2)  j^\mu(q). 
\eeq 
As long as  it is possible to isolate the s-channel production,  the form factors $f_4^Z$ and $f_5^Z$ are measurable. 
In particular, {  $f_4^Z$} can be related to experimentally observable CP asymmetries in $Z Z$ production in colliders \cite{Chang:1994cs,Grzadkowski:2016lpv}. 
On the other hand, the remaining form factors $\tilde f_i(q^2)$ in \eref{myZZZvertex} are {\em not} observable; 
in fact, they may be gauge-dependent in specific calculations.

\section{\label{sec:notation}
Complex Two-Higgs Doublet Model}

In this section we summarize the salient features of the C2HDM, 
for a review see e.g.~\cite{Gunion:1989we,Branco:2011iw}. 
The most general renormalizable scalar potential is
\begin{eqnarray}
V
&=& m_{11}^2\Phi_1^\dagger\Phi_1+m_{22}^2\Phi_2^\dagger\Phi_2
-[m_{12}^2\Phi_1^\dagger\Phi_2+{\rm h.c.}]\nonumber\\[6pt]
&&\quad +\frac{1}{2}\lambda_1(\Phi_1^\dagger\Phi_1)^2
+\frac{1}{2}\lambda_2(\Phi_2^\dagger\Phi_2)^2
+\lambda_3(\Phi_1^\dagger\Phi_1)(\Phi_2^\dagger\Phi_2)
+\lambda_4(\Phi_1^\dagger\Phi_2)(\Phi_2^\dagger\Phi_1)
\nonumber\\[6pt]
&&\quad +\left[\frac{1}{2}\lambda_5(\Phi_1^\dagger\Phi_2)^2
+\lambda_6(\Phi_1^\dagger\Phi_1)(\Phi_1^\dagger\Phi_2)
+\lambda_7(\Phi_2^\dagger\Phi_2)(\Phi_1^\dagger\Phi_2)+{\rm h.c.}\right]\,,
\label{pot}
\end{eqnarray}
where $\Phi_1$ and $\Phi_2$ are complex scalar $SU(2)_{L}$ doublets,
with vacuum expectation values (VEVs) $v_1/\sqrt{2}$ and
$v_2/\sqrt{2}$.
The parameters $m_{11}^2$, $m_{22}^2$, and $\lambda_1\ldots\lambda_4$
are real parameters,
while $m_{12}^2$ and $\lambda_5\ldots\lambda_7$ can
be complex.

In  general, both $\Phi_1$ and $\Phi_2$ can have Yukawa couplings
to all the SM fermions. However, this leads to flavour changing neutral
scalar interactions (FCNSI),
which are tightly constrained by experiment.
As a result, it is usually assumed that there is a
$\mathbb{Z}_2$ symmetry \cite{Glashow:1976nt, Paschos:1976ay},
acting on the scalars as
\be
\Phi_1 \rightarrow \Phi_1, \ \ \ \
\Phi_2 \rightarrow - \Phi_2,
\label{Z2}
\ee
with appropriate transformations on the fermions,
guaranteeing that fermions of a given charge couple exclusively
to one of the two scalar fields.

Of course,
one can perform a basis change on the scalar fields.
The couplings in the scalar potential and in the Yukawa interactions,
as well as the specific implementation of the $\mathbb{Z}_2$ symmetry,
change from one basis to the next;
but any physical observable cannot depend on such a choice.
We denote by  the ``$\mathbb{Z}_2$ basis'',
the basis in which the transformation has the specific form in
eq.~\eqref{Z2}.
For an exact $\mathbb{Z}_2$ symmetry,
$m_{12}^2$, $\lambda_6$, and $\lambda_7$ vanish.
Since the absence of $m_{12}^2$ precludes a decoupling limit
\cite{Gunion:2002zf},
one usually breaks it softly through $m_{12}^2 \neq 0$.
If $\arg(\lambda_5) = 2 \arg(m_{12}^2)$,
then we may take both couplings real,
and the potential preserves CP.
When $v_1$ and $v_2$ are also real,
there is no CP violation (explicit or spontaneous)
and the model is known as the ``real 2HDM''.
In contrast,
if $\arg(\lambda_5) \neq 2 \arg(m_{12}^2)$,
then the potential violates CP explicitly;
this is known as the ``complex 2HDM'' (C2HDM)~\cite{Ginzburg:2002wt, Khater:2003wq, ElKaffas:2007rq, ElKaffas:2006gdt, Grzadkowski:2009iz, Arhrib:2010ju, Barroso:2012wz, Fontes:2014xva}.
Here,
we will choose a basis where $v_1$ and $v_2$ are real,
without loss of generality.

It is convenient to introduce the ``Higgs basis''
\cite{Lavoura:1994fv,Botella:1994cs},
defined as the basis where only the first scalar has a VEV.
This is obtained  through the unitary transformation
\be
\left(
\begin{array}{c}
H_1\\
H_2
\end{array}
\right)
=
\left(
\begin{array}{cc}
c_\beta & s_\beta \\
- s_\beta & c_\beta
\end{array}
\right)\,
\left(
\begin{array}{c}
\Phi_1\\
\Phi_2
\end{array}
\right),
\label{changeTOHiggs}
\ee
where
$c_\beta = \cos{\beta} = v_1/v$,
$s_\beta = \sin{\beta} = v_2/v$,
and $v = \sqrt{v_1^2 + v_2^2} = (\sqrt{2} G_F)^{-1/2}$.
The doublets in the Higgs basis may be parametrized as
\begin{equation}
H_1 =
\begin{pmatrix}
-i G^+ \\
\frac{1}{\sqrt{2}} (v + h + i G^0)
\end{pmatrix}
\, ,
\ \ \ \ \
H_2 =
\begin{pmatrix}
H^+ \\
\frac{1}{\sqrt{2}} (R + i I)
\end{pmatrix}\, ,
\label{H1H2}
\end{equation}
where $G^\pm$ and $G^0$ are the Goldstone bosons which,
in the unitary gauge, are absorbed as the longitudinal components
of $W^\pm$ and $Z$,
while $H^\pm$ are the charged scalars.

The scalar potential in the Higgs basis has the form
\begin{equation}
\begin{split}
V_H =\;& Y_1 |H_1|^2 + Y_2 |H_2|^2 + (Y_3 H_1^\dagger H_2
+~\textrm{h.c.})
+ \frac{Z_1}{2} |H_1|^4 + \frac{Z_2}{2} |H_2|^4 \\
& + Z_3 |H_1|^2 |H_2|^2 + Z_4 (H_1^\dagger H_2)(H_2^\dagger H_1) \\
& + \left\{ \frac{Z_5}{2} (H_1^\dagger H_2)^2
+ (Z_6 |H_1|^2 + Z_7 |H_2|^2)(H_1^\dagger H_2)
+ \textrm{h.c.} \right\}\, ,
\label{VH}
\end{split}
\end{equation}
where we follow the notation of \cite{Bernon:2015qea}.
The parameters $Y_{1,2}$ and $Z_{1,2,3,4}$ are all real;
the others are, in general, complex.
Note that, in the Higgs basis, $H_1$ and $H_2$ are not
eigenstates of the $\mathbb{Z}_2$ symmetry and,
therefore, the cross terms proportional to $Z_6$ and $Z_7$ are in
general present.
The stationarity conditions in the Higgs basis read
\be
Y_1 = - \frac{Z_1}{2} v^2\, ,
\ \ \
Y_3 = - \frac{Z_6}{2} v^2\, .
\label{stationarity}
\ee
The last equation means that only $Z_5$, $Z_6$, and $Z_7$ are
independently complex. Thus, all sources of CP violation in the
Higgs potential must be related to the invariant quantities
$\textrm{Im}(Z_7 Z_6^\ast)$,
$\textrm{Im}(Z_7^2 Z_5^\ast)$,
and $\textrm{Im}(Z_6^2 Z_5^\ast)$ \cite{Lavoura:1994fv}.\footnote{If all
three invariants are non-vanishing, then only two are independent. But one needs
all three in order to cover also the cases in which two invariants vanish but
the third does not.}

The dictionary between the $\mathbb{Z}_2$ basis and the Higgs basis for
the quadratic terms $Y_i$ is
\begin{align}
Y_1 &= m_{11}^2 c^2_\beta + m_{22}^2 s^2_\beta
    + 2 \textrm{Re}{(m_{12}^2)} s_\beta c_\beta\, , \\
Y_2 &= m_{11}^2 s^2_\beta + m_{22}^2 c^2_\beta
    - 2 \textrm{Re}{(m_{12}^2)} s_\beta c_\beta\, , \\
Y_3 &= (m_{22}^2 - m_{11}^2) s_\beta c_\beta
    + m_{12}^2 c^2_\beta - m_{12}^{*\;2} s^2_\beta\, .
\end{align}
The generated cross-term $H_1^\dagger H_2 + \textrm{h.c.}$
(coefficient $Y_3$) can be present even if $m_{12}^2 = 0$,
unless $m_{11}^2 = m_{22}^2$ (the masses of $\Phi_1$ and $\Phi_2$ being equal).
Similarly,
for the quartic terms $Z_i$:
\begin{align}
Z_1 &= \lambda_1 c^4_\beta + \lambda_2 s^4_\beta
    + 2 \lambda_{345} s^2_\beta c^2_\beta\, , \\
Z_2 &= \lambda_1 s^4_\beta + \lambda_2 c^4_\beta
    + 2 \lambda_{345} s^2_\beta c^2_\beta\, , \\
Z_{i = 3,4} &= (\lambda_1 + \lambda_2 - 2 \lambda_{345})
    s^2_\beta c^2_\beta + \lambda_i\, , \\
Z_5 &= (\lambda_1 + \lambda_2 - 2 \lambda_{345})
    s^2_\beta c^2_\beta + \lambda_5 c^2_\beta + \lambda_5^* s^2_\beta\, , \\
Z_6 &= - s_\beta c_\beta \left[ \lambda_1 c^2_\beta
    - \lambda_2 s^2_\beta - \lambda_{345} c_{2\beta}
    - i \textrm{Im}{(\lambda_5)} \right]\, , \\
Z_7 &= - s_\beta c_\beta \left[ \lambda_1 s^2_\beta
    - \lambda_2 c^2_\beta + \lambda_{345} c_{2\beta}
    + i \textrm{Im}{(\lambda_5)} \right]\, ,
\end{align}
where $\lambda_{345}=\lambda_3 + \lambda_4 + \textrm{Re}(\lambda_5)$.
What is relevant is that not all $Z_i$ are independent,
as they satisfy the relations:
\begin{equation}
\begin{aligned}
Z_2 - Z_1 &= \frac{1 - 2 s_\beta^2}{s_\beta c_\beta}  \textrm{Re}(Z_6 + Z_7) \, , \\
Z_{345} - Z_1 &= \frac{1 - 2 s_\beta^2}{s_\beta c_\beta} \textrm{Re}(Z_6)
    - \frac{2 s_\beta c_\beta}{1 - 2 s_\beta^2} \textrm{Re}(Z_6 - Z_7) \, ,\\
\textrm{Im}(Z_6 + Z_7) &= 0\, ,\\
\textrm{Im}(Z_6 - Z_7) &= \frac{2 c_\beta s_\beta}{1 - 2 s_\beta^2} \textrm{Im}(Z_5)\, ,
\end{aligned}
\end{equation}
where $Z_{345} \equiv Z_3 + Z_4 + \textrm{Re}(Z_5)$, and the first two
equations are those relevant for the real 2HDM discussed in
\cite{Bernon:2015qea}. Using these relations we can eliminate
for example $Z_6$ and $Z_7$, and express our results in terms of the
remaining $Z_i$.
Thus,
in the C2HDM,
all CP violation invariants in the Higgs potential are proportional
to a single phase, which comes from
$\textrm{Im} (m_{12}^2 \lambda_5^\ast) $ in the original basis.

One goes from the neutral scalars $\{h, R, I\}$ written in the Higgs basis
into the neutral scalar mass basis $\{h_1, h_2, h_3\}$ through
\cite{Fontes:2014xva}
\be
\left(
\begin{array}{c}
h_1\\
h_2\\
h_3
\end{array}
\right)
=
T^T
\left(
\begin{array}{c}
h\\
R\\
I
\end{array}
\right)\, ,
\ee
where
\be
T^T =
\left(
\begin{array}{ccc}
\tilde{c}_1 c_2 & \tilde{s}_1 c_2 & s_2\\
-(\tilde{c}_1 s_2 s_3 + \tilde{s}_1 c_3)
    & \tilde{c}_1 c_3 - \tilde{s}_1 s_2 s_3  & c_2 s_3\\
- \tilde{c}_1 s_2 c_3 + \tilde{s}_1 s_3
    & -(\tilde{c}_1 s_3 + \tilde{s}_1 s_2 c_3) & c_2 c_3
\end{array}
\right)
\label{matrixT}
\ee
and $s_i = \sin{\alpha_i}$ and
$c_i = \cos{\alpha_i}$ ($i = 2, 3$).
Similarly,
$\tilde{s}_1 = \sin{\tilde{\alpha}_1}$
and
$\tilde{c}_1 = \cos{\tilde{\alpha}_1}$,
where $\tilde{\alpha}_1=\alpha_1 - \beta$.
We have defined $T$ to agree with the definition in
ref.~\cite{Branco:2011iw}.
In the C2HDM one usually defines a matrix $R$
such that
\be
T^T = R R_H
=
\left(
\begin{array}{ccc}
c_1 c_2 & s_1 c_2 & s_2\\
-(c_1 s_2 s_3 + s_1 c_3) & c_1 c_3 - s_1 s_2 s_3  & c_2 s_3\\
- c_1 s_2 c_3 + s_1 s_3 & -(c_1 s_3 + s_1 s_2 c_3) & c_2 c_3
\end{array}
\right)
\left(
\begin{array}{ccc}
c_{\beta} & - s_{\beta} & 0\\
s_{\beta} & c_{\beta} & 0\\
0 & 0 & 1
\end{array}
\right).
\ee
The angles
$\alpha_1$, $\alpha_2$,
and $\alpha_3$ were introduced in \cite{ElKaffas:2007rq},
and, without loss of generality,
may be restricted to
\be
- \pi/2 < \alpha_1 \leq \pi/2,
\hspace{5ex}
- \pi/2 < \alpha_2 \leq \pi/2,
\hspace{5ex}
0 \leq \alpha_3 \leq \pi/2.
\label{range_alpha}
\ee
The real 2HDM may be obtained by setting
$s_2 = s_3 = 0$,
and the usual $\alpha =\alpha_1 - \pi/2$.

In the Higgs basis $h$ is the only scalar field that has a coupling to two gauge bosons,
and it coincides with the SM one $g^\textrm{sm}_{h V V}$.
Thus,
\be
g_{h_k V V}
=
g^\textrm{sm}_{h V V}\,
T_{1k}\, .
\label{hkVV}
\ee
In the C2HDM,
\be
T_{1k} = c_\beta R_{k1}+ s_\beta R_{k2}\, ,
\label{T1i}
\ee
and the coupling of the lightest Higgs to gauge bosons is given by
\be
g_{h_1 V V} = g^\textrm{sm}_{h V V}\, c_2\, \cos{(\alpha_1 - \beta)},
\ee
which reduces in the real 2HDM to
$g_{h_1 V V} = g^\textrm{sm}_{hV V}\, \sin{(\beta - \alpha)}$.
Notice that, since the matrix $T$ is orthogonal,
eq.~\eqref{hkVV} implies that
\be
\label{eq:sumrule}
\sum_k \left| g_{h_k V V} \right|^2 = \left| g^\textrm{sm}_{h V V} \right|^2
\ee
and the coupling of each scalar mass eigenstate with two vector bosons
must be smaller than the corresponding coupling in the SM.
This property generalizes to any multi Higgs doublet model,
so that a value well above the SM would exclude the SM and
also all such models.
Conversely,
since the measurements are consistent with a coupling of the 125 GeV
scalar with two gauge bosons very close to the SM value, then
the mixing angles in $T$ must be such that this scalar almost coincides
with the $h$ in the Higgs basis.
This translates into the so-called \textit{alignment limit} of
\be
s_2 \rightarrow 0
\ \ \
\textrm{and}
\ \ \
\sin{(\alpha_1 - \beta)} \rightarrow 0
\ \ \ \ \ \ \ \textrm{(C2HDM)}\, ,
\ee
and
\be
\cos{(\beta - \alpha)} \rightarrow 0
\ \ \ \ \ \ \ \textrm{(real 2HDM)}\, ,
\ee
in the C2HDM and real 2HDM, respectively.

\section{\label{sec:ZZZ_C2HDM}CP-violating $ZZZ$ vertex in C2HDM}

We  turn to the calculation of one-loop contributions to the $ZZZ$ vertex in the C2HDM.
The goal is to determine the CP-violating form factor $f_4^Z(q^2)$ defined by \eref{myZZZvertex} (the other physical form factor $f_5^Z(q^2)$ vanishes at one loop). 
To that end, we can neglect all the Lorentz structures that are not of the form
$\eta^{\mu \alpha} p_1^\beta $  or  $ \eta^{\mu\beta} p_2^\alpha$. 
A good consistency check is to verify that these two Lorentz structures have the same
coefficient, as the result should be invariant for the exchange $(p_1,\alpha)\leftrightarrow (p_2,\beta)$.
We perform the calculation in a general $R_\xi$ gauge, and verify gauge invariance at the end of the calculation. 
We express the results in terms of the Passarino-Veltman (PV) functions~\cite{Passarino:1978jh}, following the \texttt{LoopTools} conventions~\cite{Hahn:1998yk}. 
To evaluate the loop integrals we use the \texttt{Mathematica} packages \texttt{FeynCalc}~\cite{Mertig:1990an}, and we cross-checked the result with \texttt{Package-X}~\cite{Patel:2015tea}. 
Our final result disagrees with the previous literature~\cite{Grzadkowski:2016lpv},
therefore we will present in some detail the intermediate steps of our calculation.

\subsection{Couplings and Propagators}

For our calculation, we need the following vertices~\cite{WebPageC2HDM},
\begin{align}
[h_i, h_j, Z^{\mu}] =& \frac{g}{2 c_W} \left(p_i-p_j\right)^\mu
\epsilon_{ijk}\, x_k\, ,
\label{eq:3}
\\[+2mm]
[ Z^\mu,G^0,h_i] = &   \frac{g}{2 c_W} \left(p_i-p_0\right)^\mu \, x_i\, ,
\label{eq:4}
\\[+2mm]
[h_i, Z^\mu, Z^\nu] = & i\, \frac{g}{c_W}\, m_Z\, g^{\mu\nu} \, x_i\, ,
\label{eq:5}
\end{align}
where $c_W= \cos\theta_W$,  all momenta are incoming, and the $i$ of the Feynman rules is
already included.\footnote{%
Note that we use the convention for the gauge couplings where the covariant derivatives are written as 
$D_\mu = \partial_\mu + i g A_\mu$. 
If the opposite convention ($D_\mu = \partial_\mu - i g A_\mu$) were used, then the sign of the vertices $[h_i, h_j, Z^{\mu}]$ and $[ Z^\mu,G^0,h_i]$ would be flipped  
and the $Z^3$ form factor calculated below would pick up an overall minus sign.  
When the form factor is included for example into the  $f \bar f \rightarrow Z^* \rightarrow Z Z$  amplitude,
this sign cancels with the corresponding sign choice for the [$f, \bar f, Z$] vertex; only the product $[f, \bar f, Z] f_4^Z$ has physical meaning. 
}
The coefficient $x_i$ above is related to the C2HDM parameters as
\begin{equation}
  \label{eq:2}
  x_i\equiv T_{1i} = \left[ c_\beta R_{i1} + s_\beta  R_{i2}   \right] \, .
\end{equation}
That is, $x_i$ coincide with the $T_{1i}$ in eq.~\eqref{T1i}.
As we are doing the calculation in a general $R_\xi$ gauge,
we also need the propagators for the Goldstone $G^0$ and
the $Z$ in this gauge~\cite{Romao:2012pq},
\begin{align}
[G^0, G^0] =& \frac{i}{p^2- \xi m_Z^2 + i\, \epsilon}\, ,
\\[+2mm]
[Z^\mu,Z^\nu] =& -i\, \frac{1}{k^2 -m_Z^2 + i\, \epsilon} \left[
g^{\mu\nu} -  (1-\xi) \frac{k^\mu k^\nu}{k^2 -\xi m_Z^2} \right]\, .
\end{align}

\subsection{Diagrams with $h_i, h_j, h_k$}

We start with the diagrams containing only Higgs bosons in the internal lines, as shown in \fref{zzzhhh}.
Because of the coupling structure in \eref{3}, all the three scalars have to be different. 
\begin{figure}[htb]
  \centering
  \includegraphics[scale=1]{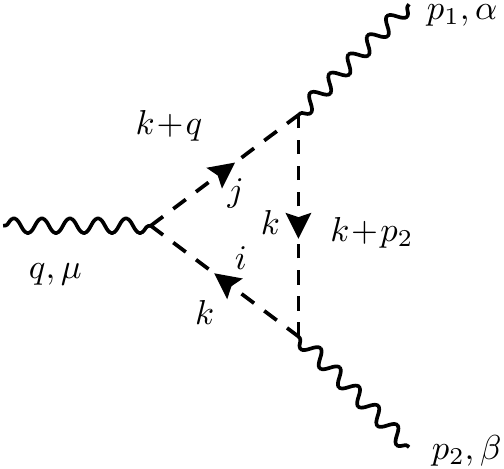}
  \caption{Contribution to the $Z^3$ vertex $\Gamma_{\mu \alpha \beta}$ with three Higgs $h_i,h_j,h_k$ in the loop.}
  \label{fig:zzzhhh}
\end{figure}
We get the same result as in Ref.~\cite{Grzadkowski:2016lpv},
\begin{equation}
  \label{eq:6}
  e \frac{q^2-m_Z^2}{m_Z^2} \, f_4^{Z,hhh} = - \frac{8}{16\pi^2}
  \left(\frac{g}{2 c_W}\right)^3  x_1 x_2 x_3
  \sum_{i,j,k} \epsilon_{ijk}C_{001}(q^2,m_Z^2,m_Z^2,m_i^2,m_j^2,m_k^2)\, . 
\end{equation}
%

\subsection{Diagrams with $h_i, h_j, G^0$}

We consider now the diagram with one Goldstone boson in one of the
internal lines (there are no diagrams with either two or three
Goldstone bosons), as shown in~\fref{zzzhhg}.
\begin{figure}[htb]
  \centering
  \includegraphics[scale=1]{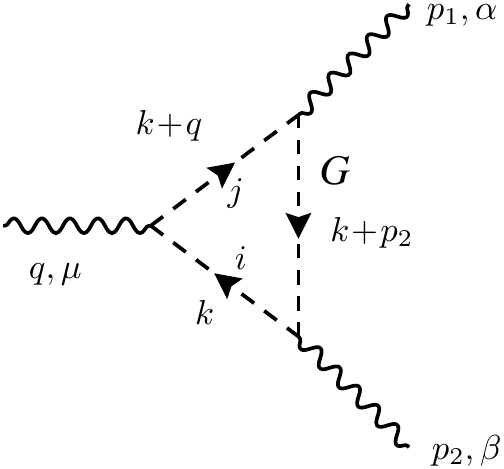}
  \caption{Contribution to  the $Z^3$ vertex $\Gamma_{\mu \alpha \beta}$  with  $h_i,h_j,G^0$ in the loop.}
  \label{fig:zzzhhg}
\end{figure}
There are two more diagrams with the $G^0$ in the other internal
lines. Each of them will have all the possible combinations of
$h_i,h_j$. As before, due to the coupling structure in
eq.~(\ref{eq:3}) and eq.~(\ref{eq:4}), we must have $i\not= j$ in all
possible combinations.
In the $R_\xi$ gauge,
we get the result,
\begin{align}
\label{eq:7}
  e \frac{q^2-m_Z^2}{m_Z^2} \, f_4^{Z,hhG} =&  \frac{8}{16\pi^2}
  \left(\frac{g}{2 c_W}\right)^3  x_1 x_2 x_3
  \sum_{i,j,k} \epsilon_{ijk}  \left[
C_{001}(q^2,m_Z^2,m_Z^2,m_i^2,m_j^2,\xi m_Z^2) \right.\nonumber\\[+2mm]
+&\left.
C_{001}(q^2,m_Z^2,m_Z^2,\xi m_Z^2,m_j^2,m_k^2)  +
C_{001}(q^2,m_Z^2,m_Z^2,m_i^2,\xi m_Z^2,m_k^2)
\right]\, ,
\end{align}
which agrees with Ref.~\cite{Grzadkowski:2016lpv} in the Feynman gauge limit $\xi=1$.

\subsection{Diagrams with $h_i, h_j, Z$}

Finally, we evaluate the contribution from the diagrams with one $Z$ boson
in an internal line (again there are no diagrams with two or three $Z$
bosons in internal lines), as shown in~\fref{zzzhhZ}.
\begin{figure}[htb]
  \centering
  \includegraphics[scale=1]{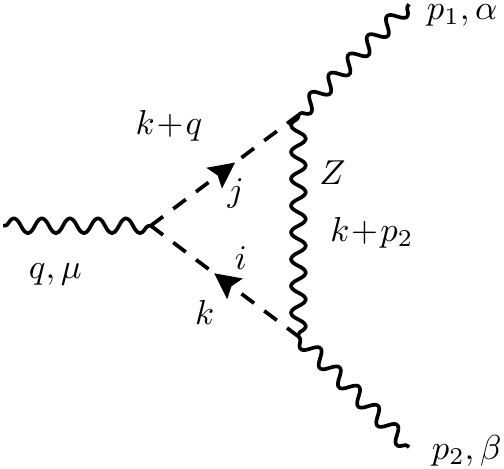}
  \caption{Contribution to the $Z^3$ vertex $\Gamma_{\mu \alpha \beta}$  with $h_i,h_j,Z^0$ in the loop.}
  \label{fig:zzzhhZ}
\end{figure}
We get
\begin{align}
\label{eq:8}
  e \frac{q^2-m_Z^2}{m_Z^2} \, f_4^{Z,hhZ} =&  \frac{8}{16\pi^2}
  \left(\frac{g}{2 c_W}\right)^3  x_1 x_2 x_3
  \sum_{i,j,k} \epsilon_{ijk}  \left[
C_{001}(q^2,m_Z^2,m_Z^2,m_i^2,m_j^2, m_Z^2) \right.\nonumber\\[+2mm]
&+\left.
C_{001}(q^2,m_Z^2,m_Z^2, m_Z^2,m_j^2,m_k^2)  +
C_{001}(q^2,m_Z^2,m_Z^2,m_i^2, m_Z^2,m_k^2)
\right]\nonumber\\[+2mm]
& - \frac{8}{16\pi^2}
  \left(\frac{g}{2 c_W}\right)^3  x_1 x_2 x_3
  \sum_{i,j,k} \epsilon_{ijk}  \left[
C_{001}(q^2,m_Z^2,m_Z^2,m_i^2,m_j^2,\xi m_Z^2) \right.\nonumber\\[+2mm]
&+\left.
C_{001}(q^2,m_Z^2,m_Z^2,\xi m_Z^2,m_j^2,m_k^2)  +
C_{001}(q^2,m_Z^2,m_Z^2,m_i^2,\xi m_Z^2,m_k^2)
\right]\nonumber\\[+2mm]
&- \frac{8}{16\pi^2}
  \left(\frac{g}{2 c_W}\right)^3  x_1 x_2 x_3\, m_Z^2\,
  \sum_{i,j,k} \epsilon_{ijk}
C_{1}(q^2,m_Z^2,m_Z^2,m_i^2,m_Z^2,m_k^2)\, .
\end{align}
In the limit $\xi=1$ this result differs in the overall sign from that in  Ref.~\cite{Grzadkowski:2016lpv}. 

\subsection{Final Result}

\begin{figure}
\bc 
\includegraphics[width=0.45\textwidth]{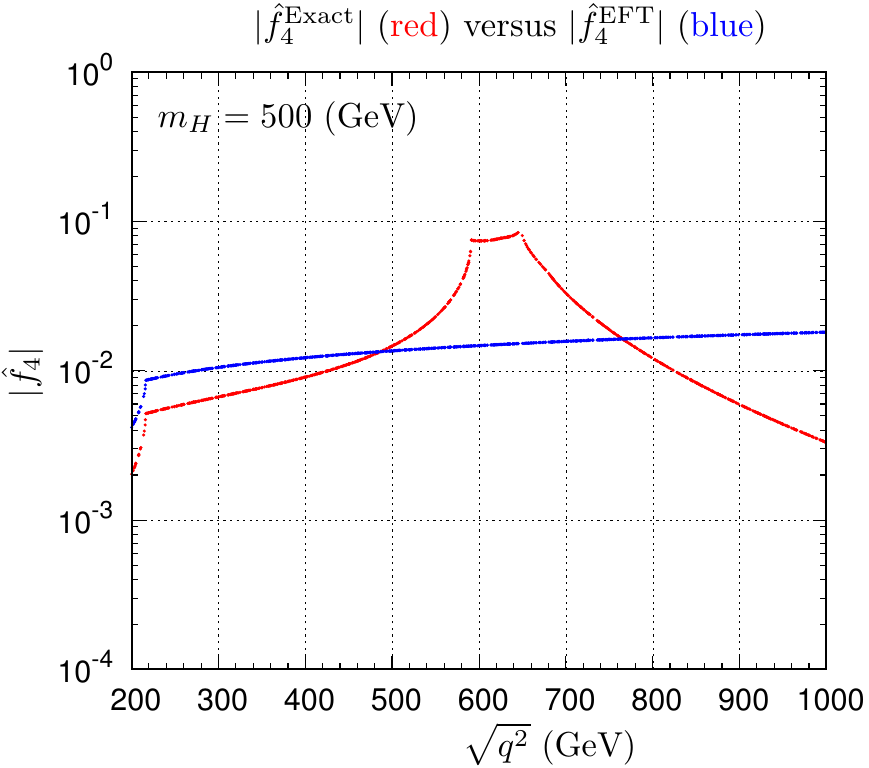}
\includegraphics[width=0.45\textwidth]{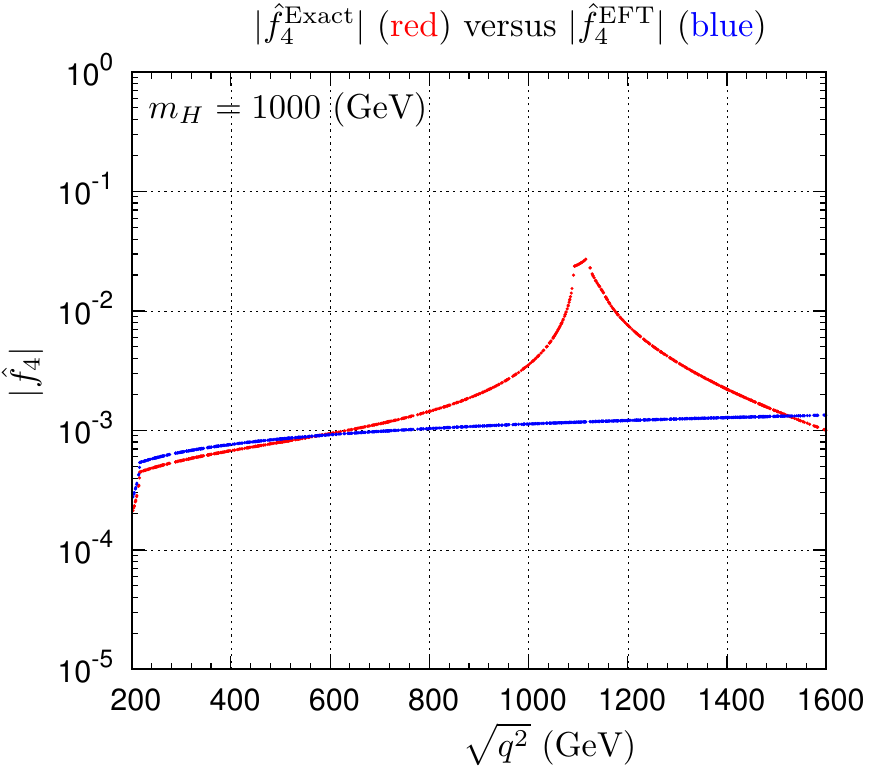}
\includegraphics[width=0.45\textwidth]{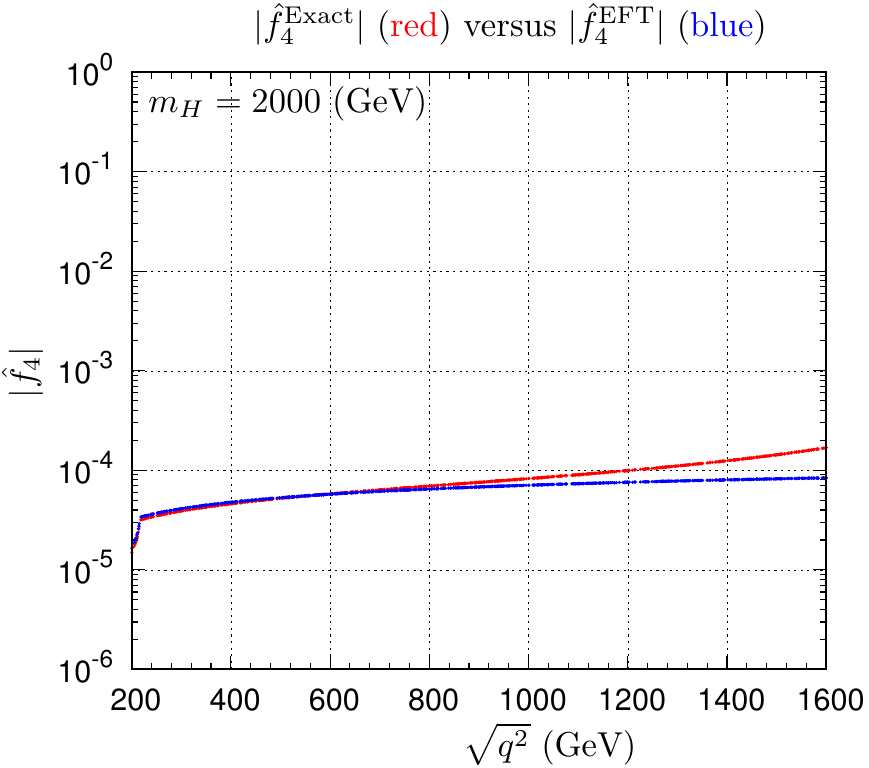}
 \ec 
 \caption{
 \label{fig:f4z}
Red: the normalized form factor $\hat f_4^Z(q^2)$ as defined in \eref{9}.  
We show the dependence on $\sqrt{q^2}$ for 3 different values $m_H$ of the second neutral Higgs mass. 
The heaviest neutral Higgs mass is  assumed to be  $\sqrt{m_H^2 + v^2}$.   
Blue: the same observable calculated in the matched  SMEFT (cf.\eref{ZZZ_FzzzEFT} in \sref{eft2hdm}). 
  }
 \end{figure}

Summing the different contributions, $f_4^Z =  f_4^{Z,hhh} +  f_4^{Z,hhZ} + f_4^{Z,hhG}$, 
we find that the $\xi$ dependent parts of Eqs.~(\ref{eq:7}) and (\ref{eq:8})
cancel out, ensuring gauge invariance of the final result. 
Also the antisymmetry of each term in eqs.~(\ref{eq:6}), (\ref{eq:7}) and
(\ref{eq:8}) implies that the divergences originating from the PV function $C_{001}$ cancel and the final result is
finite. 
All in all, the CP violating $Z^3$ form factor expressed by the PV functions takes the form 
\ba
&&
  e \frac{q^2-m_Z^2}{m_Z^2} \, f_4(q^2) 
  \left[ \frac{1}{16\pi^2}
  \left(\frac{g}{c_W}\right)^3 x_1 x_2 x_3 
  \right]^{-1}\, \equiv \hat f_4^Z =
\nonumber\\[+2mm]
&&
\hspace{6ex}
\sum_{i,j,k} \epsilon_{ijk} \left[ -\, C_{001}(q^2,m_Z^2,m_Z^2,m_i^2,m_j^2,m_k^2)
+\, C_{001}(q^2,m_Z^2,m_Z^2,m_i^2,m_j^2, m_Z^2)
\right.
\nonumber\\[+2mm]
&&
\hspace{14ex}
+\, C_{001}(q^2,m_Z^2,m_Z^2, m_Z^2,m_j^2,m_k^2)  +
C_{001}(q^2,m_Z^2,m_Z^2,m_i^2, m_Z^2,m_k^2)
\nonumber\\[+2mm]
&&
\hspace{14ex}
\left.
-\, m_Z^2\,
C_{1}(q^2,m_Z^2,m_Z^2,m_i^2,m_Z^2,m_k^2)
\right]\, .
\label{eq:9}
\ea
The dependence of the form factor on $q^2$ is illustrated in \fref{f4z} for several choices of the heavy scalar spectrum. 
The order of magnitude of $|f_4^Z|$ that can be achieved in the realistic parameter space of the C2HDM is shown in  \fref{f4zscatter}, reaching the values of order $10^{-5}$.   
For comparison, the recent ATLAS~\cite{Aaboud:2017rwm} and CMS~\cite{Sirunyan:2017zjc} analyses of $ZZ$ production at the LHC set upper bounds on  $|f_4^Z|$ (assumed real) on the order of $10^{-3}$. 
When considering a generic framework beyond the SM, one must check whether effects other than
$f_4^Z$ may contribute to the actual experimental observable being measured (and from which
$f_4^Z$ is inferred).
{  For example, one can see from Fig.~1 in Ref.~\cite{Sirunyan:2017zjc} that there
is a contribution from $h_{125} \rightarrow ZZ$
to the four lepton events from which $f_4^Z$
is extracted. For the SM Higgs this is not a problem,
since this is merely an order $5$\% contribution to the cross section and,
moreover, the measurement of $f_4^Z$ is made by requiring in addition
that each $Z$ in the final state has a mass in the range  60-120 GeV.
But it could be a concern if a heavier Higgs were to decay into
$ZZ$, competing with the signal from the $ZZZ$ vertex.
This problem is mitigated in the C2HDM because of a combination of two facts.
First, we know from the $h_{125} \rightarrow ZZ$
measurements that the corresponding coupling in the C2HDM lies
very close to the SM value (the so-called alignment limit).
Second, the sum rule in \eref{sumrule} guarantees that
any heavier scalar will have a very small coupling to
$ZZ$. Nevertheless, once statistics improve at LHC, a precise
constraint on $f_4^Z$ can best be achieved by a detailed
simulation of the C2HDM within the experimental analysis of
the collaborations, which is beyond the scope of this work.}
Our results for the maximum of $|f_4^Z|$ are slightly below those reported in Ref.~\cite{Grzadkowski:2016lpv}. This is mainly due to the effect of including in our scan the bound on the electron EDM~\cite{Baron:2013eja}. 
The sign difference that we have found does not affect much the absolute value, because the diagram where it occurs is typically the dominant one (in the gauge $\xi = 1$)~\cite{Grzadkowski:2016lpv}.
 
\begin{figure}
\bc 
\includegraphics[width=0.45\textwidth]{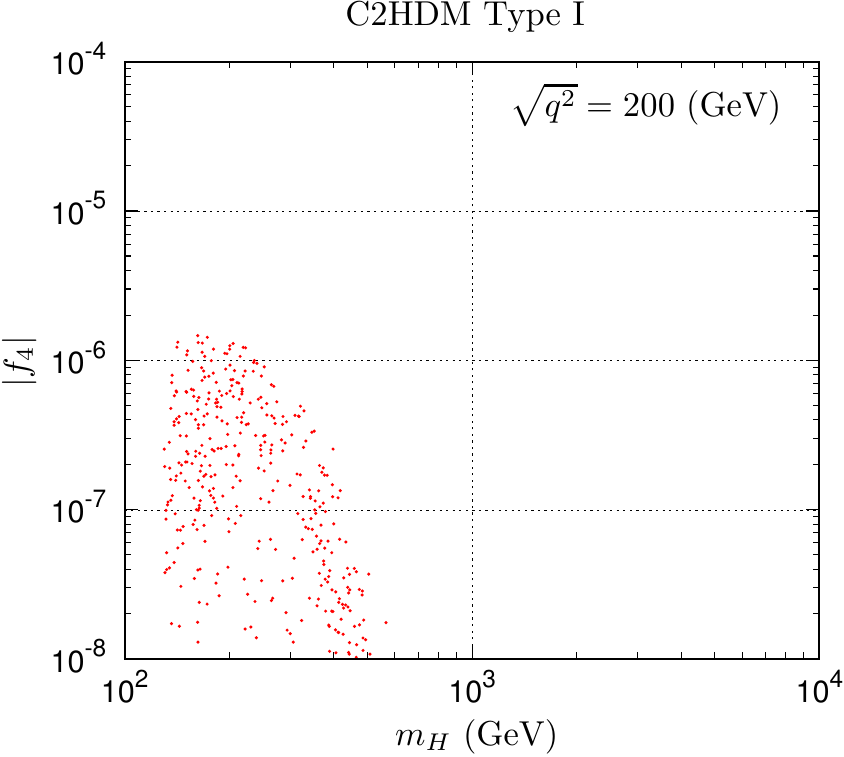}
\includegraphics[width=0.45\textwidth]{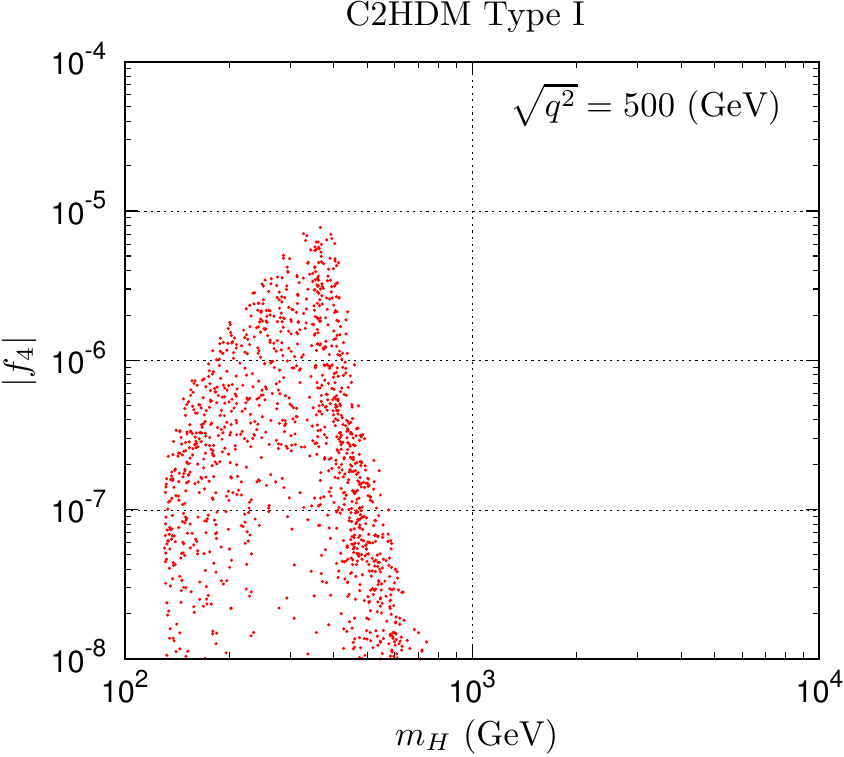}
 \ec 
 \caption{
 \label{fig:f4zscatter}
 Scatter plots showing the absolute value of the CP violating form factor $f_4^Z(q^2)$ for two values of  $\sqrt{q^2}$ for points in the parameter space of the type-1 C2HDM  satisfying theoretical (unitarity, bounded from below) and experimental (LHC Higgs, electric dipole moments, and electroweak precision measurements) constraints.} 
 \end{figure}

For future reference, we also give the final form of the $Z^3$ vertex before evaluating  the loop integrals:   
 \bea 
 \label{eq:ZZZ_GammaFinal} 
i \Gamma_{\mu \alpha \beta}  &=  &
- i  \left ( g \over c_W \right ) ^3 (x_1 x_2 x_3)  \int {d^4 k \over (2\pi)^4}  \eps_{ijk} \left \{ 
 {k_\mu  k_\alpha k_\beta   \over ( (k-p_1)^2 - m_i^2)((k+ p_2)^2 - m_j^2)  (k^2 -  m_k^2 )}
 \right . \nnl   &-&   \left.
  { k_\mu  k_\alpha k_\beta \over  ( (k-p_1)^2 - m_j^2)((k+ p_2)^2 - m_k^2) (k^2 -  m_Z^2 )} 
\right . \nnl   &-&   \left.
 { k_\mu k_\alpha  k_\beta   \over ((k-p_1)^2 - m_Z^2 ) ( (k+p_2)^2 - m_j^2)  (k^2 - m_k^2)} 
  \right . \nnl   &-&   \left.
{  k_\mu k_\alpha k_\beta  \over ( (k-p_1)^2 - m_k^2) ((k+p_2)^2 -  m_Z^2 ) (k^2 - m_j^2) } 
 \right . \nnl   &+&   \left.
 {m_Z^2  k_\beta \eta_{\mu \alpha}  \over  ((k-p_1)^2 - m_Z^2 ) ( (k+p_2)^2 - m_j^2) (k^2 - m_k^2)} 
\right . \nnl   &-&   \left.
 {m_Z^2  k_\alpha  \eta_{\mu \beta}  \over ( (k-p_1)^2 - m_j^2) ((k+p_2)^2 -  m_Z^2 ) (k^2 - m_k^2) }
 \right \}   + ({\rm ILS }), 
\eea 
where ${\rm ILS}$ stands for (in general divergent and gauge dependent) irrelevant Lorentz structures that do not contribute to the observable form factors.
In this form the vertex is manifestly symmetric under interchanging $p_1 \leftrightarrow p_2$, $\alpha \leftrightarrow \beta$. 
Performing the momentum integral and extracting from the coefficient of the  tensor structure 
$\eta_{\mu \alpha} p_{1,\beta}  + \eta_{\mu \beta} p_{2,\alpha}$, cf.\eref{myZZZvertex}, one obtains the result in \eref{9}.

\section{CP-violating $ZZZ$ vertex in SMEFT
\label{sec:eft2hdm}  
}

In this section we discuss how the CP-violating $ZZZ$ vertex arises in the low-energy EFT where the heavy non-SM scalars of the C2HDM are integrated out. 
We  denote $m_1 = m_h  = 125$~GeV, 
$m_2 = m_H$, $m_3 = \sqrt{m_H^2 +  \delta^2}$ with $\delta \sim v$, 
and we are interested in the decoupling limit $m_H  \gg m_h$. 
In such a case,  only the SM degrees of freedom are available at the energies $E \sim v \ll m_H$. 
In this regime the dynamics is described by the SMEFT, 
with the SM Lagrangian  augmented by higher-dimensional operators. 
At the level of dimension-6 operators the matching of the SMEFT Lagrangian to the 2HDM UV completion was discussed e.g. in Refs.~\cite{Perez:1995dc,Henning:2014wua,Gorbahn:2015gxa,Brehmer:2015rna,Egana-Ugrinovic:2015vgy,Belusca-Maito:2016dqe,Karmakar:2017yek}.    
However, within the EFT framework studied in these references, there is no source of CP violation contributing to the $ZZZ$ vertex. 
Below we will identify the higher-dimensional CP-violating operator and discuss how  the $ZZZ$ vertex is generated.

The first step toward this goal is to expand the form factor $f_4^Z(q^2)$ in powers of $1/m_H$. 
In principle, one could expand the result in \eref{8} using the known expressions for the PV functions. 
In practice, this path  is difficult due to a complicated form and non-analytic behavior of the PV functions involved.  
Instead, we find it easier to apply 
the method of regions~\cite{Beneke:1997zp}.
A loop integral containing two disparate mass scales $m_{\rm light} \ll m_{\rm heavy}$ can be calculated by A) expanding the integrand for $k \sim m_{\rm light}$ and performing the integral, 
B) expanding the integrand for $k \sim m_{\rm heavy}$ and performing the integral,
and then adding these two contributions together.  
An important point here is that both A) and B) have a clear counterpart on the EFT side where the scales $m_{\rm heavy}$ are integrated out. 
Namely,  A) corresponds to {\em 1-loop} Feynman diagrams with the light particles in the loop and an insertion of tree-level-generated effective operators, 
while B) corresponds to {\em tree-level}  diagrams with an insertion of operators whose Wilson coefficients are suppressed by a loop factor. 
Applying the method of regions to the integrals in \eref{ZZZ_GammaFinal} we find that the leading contributions to the sum of the integrals are $\cO(m_H^{-4})$ and come from the diagrams with one heavy scalar and two SM particles  in the loop ($h$, $Z$, or the corresponding Goldstone boson).  
Moreover, we find that it is the soft region A), $k \ll m_H$, which dominates. 
Other diagrams and integration regions contribute only at $\cO(m_H^{-6})$  or higher. 
This immediately tells us that, in the EFT for the 2HDM, the CP-violating $ZZZ$ vertex is generated at one loop via diagrams with $h$, $Z$ in the loop and an insertion of a tree-level-generated effective operator. 

In \aref{mr} we give the details of the method of regions applied to \eref{ZZZ_GammaFinal}, albeit for the sake of brevity we work there in the simplified limit $m_h \to m_Z$. 
Here we write down  the leading contribution to  the CP-violating form factor $f_4^Z$ for a general $m_h$, valid for $m_h \ll m_H$ and $q^2 \ll m_H^2$:  
\bea
\label{eq:ZZZ_FzzzEFT}
& e f_{4}^Z(q^2) \approx &  
 {\delta^2 x_1 x_2 x_3 \over m_H^4 } \left ( g \over c_W \right )^3  { 1 \over 384 \pi^2 m_Z^4 q^6 (q^2 - m_Z^2) } \left \{ 
  \right . \nnl   &+ &   \left.
2 m_h^2 m_Z^2 q^6 \left(m_h^4-5 m_h^2 m_Z^2+10 m_Z^4\right) {\rm DiscB} \left(m_Z^2,m_h,m_Z \right)
  \right . \nnl   &- &   \left.
2 m_Z^6 q^2 {\rm DiscB}  \left(q^2,m_h,m_Z\right) \times 
 \right . \nnl   & &   \left.
 \left[ m_h^6-m_h^4 \left(3 m_Z^2+2 q^2\right)+m_h^2 \left(3 m_Z^4+6 m_Z^2 q^2+q^4\right)-m_Z^2 \left(m_Z^4+4 m_Z^2 q^2-5 q^4\right)\right ]  
    \right . \nnl   &+ &   \left.
m_Z^2 q^2 \left(m_Z^2-m_h^2\right)    \left(m_Z^2-q^2\right) \left(2 m_h^4 \left(m_Z^2+q^2\right)-m_h^2 \left(4 m_Z^4+9 m_Z^2 q^2\right)+2 m_Z^6+9 m_Z^4 q^2\right)
    \right . \nnl   &+ &   \left.
 \left [ m_h^8 \left(m_Z^4+m_Z^2 q^2+q^4\right)
 -m_h^6 \left(4 m_Z^6+7 m_Z^4 q^2+7 m_Z^2 q^4\right)
 +3 m_h^4 \left(2 m_Z^8+5 m_Z^6 q^2+6 m_Z^4 q^4\right)
     \right . \right .  \nnl   & &   \left. \left . 
 -m_h^2 m_Z^6 \left(4 m_Z^4+13 m_Z^2 q^2+13 q^4\right)
 +m_Z^8 \left(m_Z^4+4 m_Z^2 q^2-5 q^4\right)\right ]  
 \left(m_Z^2 - q^2 \right)     \log \left(\frac{m_h^2}{m_Z^2}\right)
\right \},
\nnl 
\eea 
where we introduced the DiscB function as defined in~\cite{Patel:2015tea}: 
\bea
\label{eq:ZZZ_discb2}
 {\rm DiscB}(p^2,m_1,m_2)  & = &  \lambda(p^2,m_1,m_2) 
 \log \left ( m_1^2 + m_2^2  - p^2  + p^2 \lambda(p^2,m_1,m_2) \over 2 m_1 m_2 \right ), 
 \nnl 
 \lambda(p^2,m_1,m_2) &= &   \sqrt{1 -  {2 (m_1^2 + m_2^2) \over p^2}  + {(m_1^2 - m_2^2)^2 \over p^4} }. 
\eea  
We have checked numerically that \eref{ZZZ_FzzzEFT} correctly reproduces the $f_4^Z$ form factor in the C2HDM in the decoupling limit.
This is illustrated in \fref{f4zeft} where, as long as  $q^2 \ll m_H^2$, both the real and imaginary parts of the two results converge as we increase $m_H$.

\begin{figure}
\bc 
\includegraphics[width=0.45\textwidth]{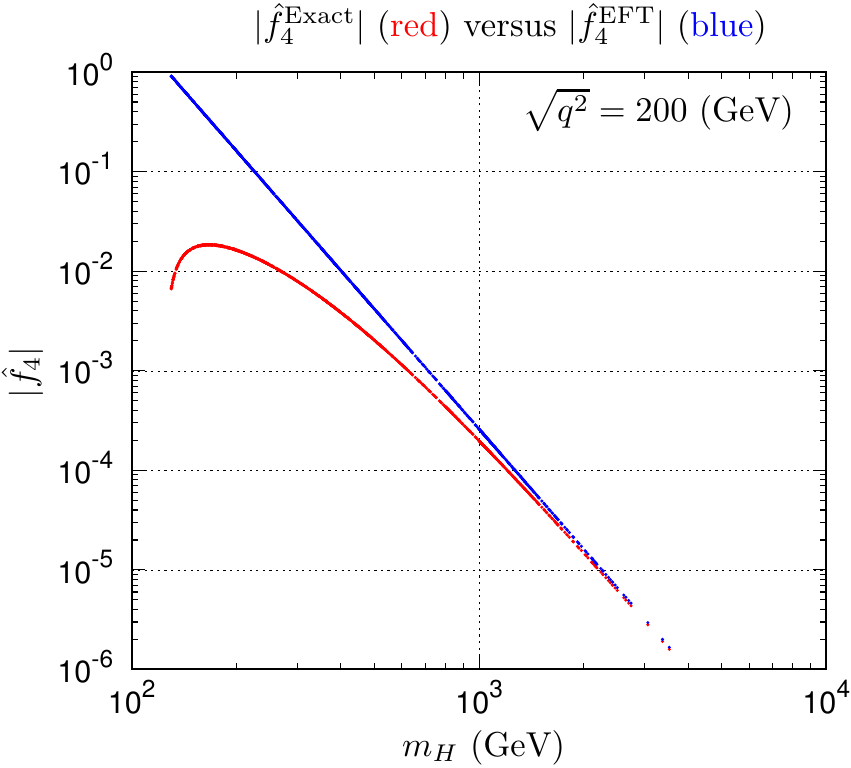}
\includegraphics[width=0.45\textwidth]{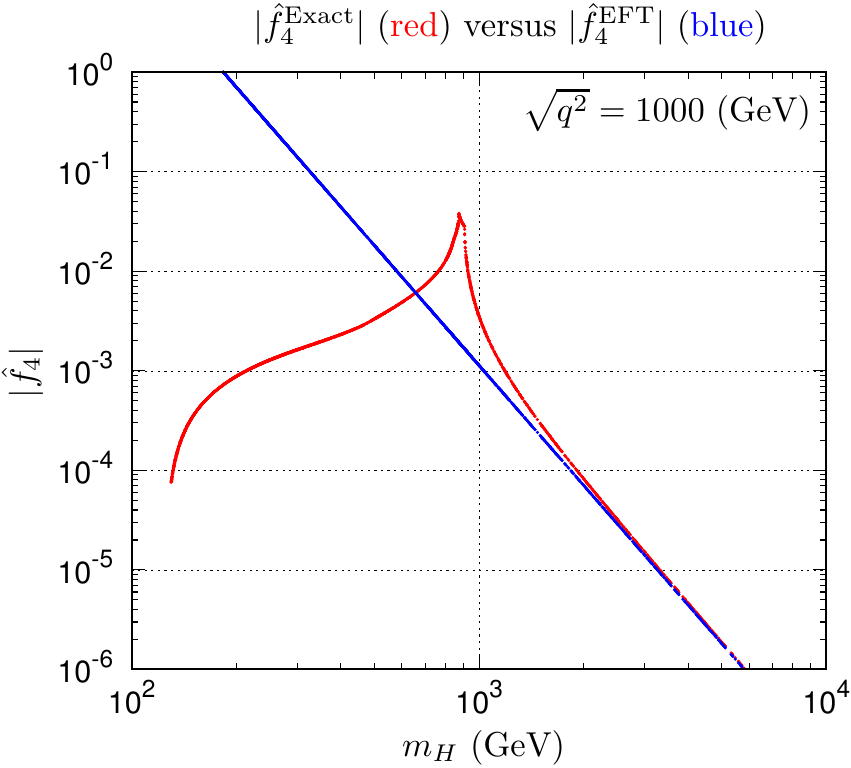}
\includegraphics[width=0.45\textwidth]{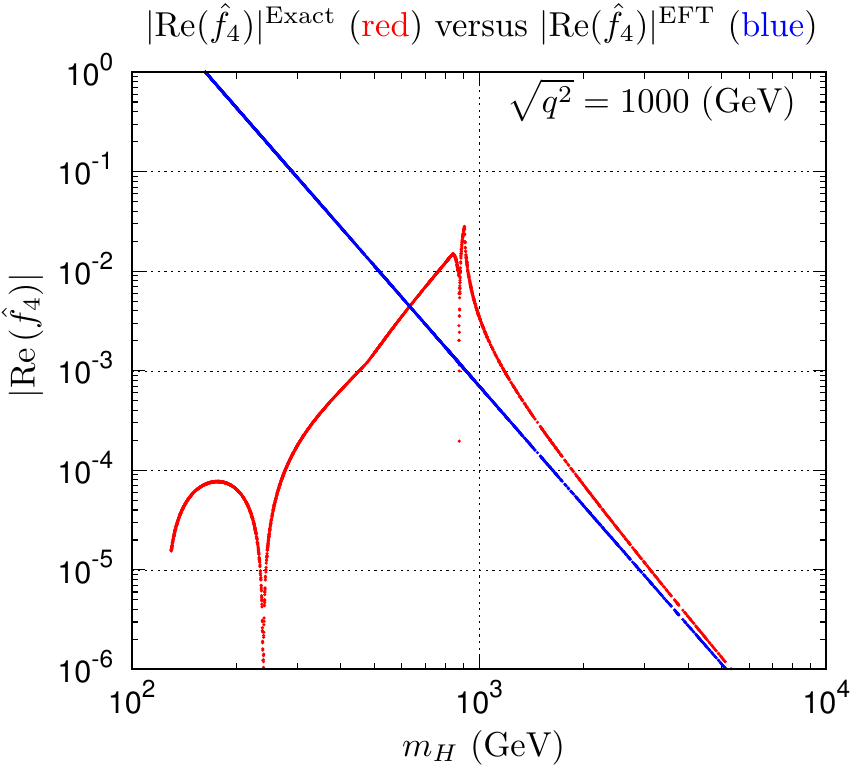}
\includegraphics[width=0.45\textwidth]{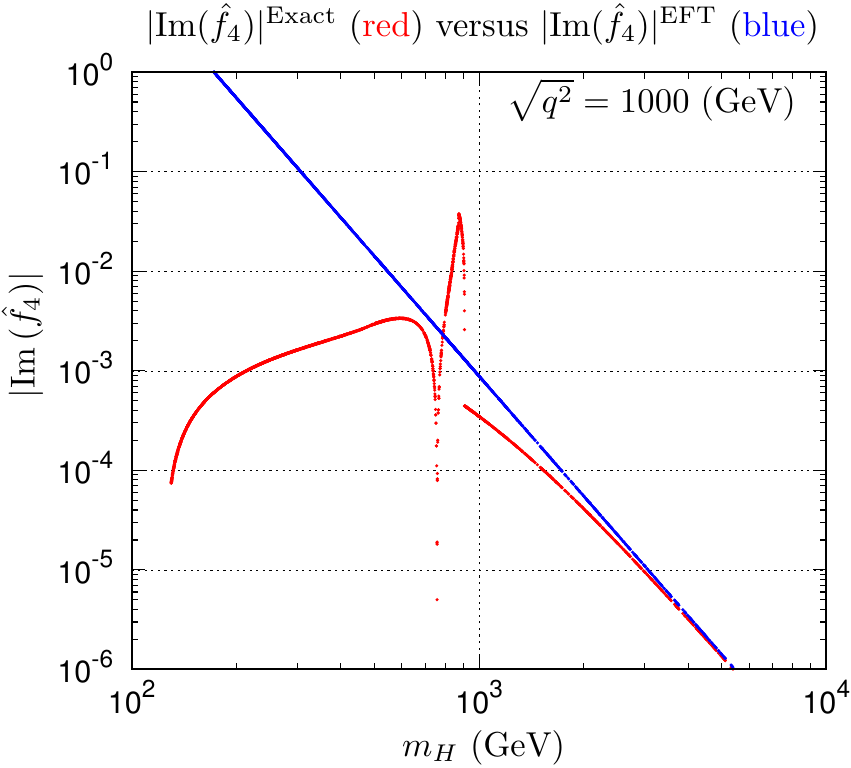}
 \ec 
 \caption{
 \label{fig:f4zeft}
 Red: the normalized form factor $\hat f_4^Z(q^2)$ as defined in \eref{9}.  
 We show the dependence on  the second neutral Higgs mass $m_H$ for two values of $\sqrt{q^2}$. 
 For $\sqrt{q^2} = 1$~TeV we also show separately the real and  imaginary parts (for $\sqrt{q^2} = 200$~GeV the form factor is purely real). 
The heaviest neutral Higgs mass is assumed to be  $\sqrt{m_H^2 + v^2}$.   
Blue: the same observable calculated in the matched SMEFT (cf.\eref{ZZZ_FzzzEFT}). 
  }
 \end{figure}

We find that the CP violating $Z^3$ form factor is strongly suppressed in the decoupling limit. 
First, the momentum integration brings the suppression factor $1/m_H^4$ in \eref{ZZZ_FzzzEFT}, which is stronger than the naive estimate from dimensional analysis  due to cancellations between diagrams with $h_2$ and $h_3$. 
Moreover, in the decoupling limit the mixing angles between the Higgs scalars are also suppressed: 
\beq
\label{eq:ZZZ_jarlskogdecoupling}
 \delta^2  x_1 x_2 x_3  \approx 
{v^6 \over 2 m_H^4} \im [Z_5^* Z_6^2] . 
\eeq  
All in all, we find that $f_4^Z \sim {1 \over (16 \pi^2) m_H^8}$ in the decoupling limit.  
This tells us that, in the SMEFT matched to C2HDM at one loop, the CP violating $ZZZ$ vertex arises from a dimension-12 operator!  
Note that  $f_4^Z(q^2)$ has an imaginary part  for $q^2 > (m_Z + m_h)^2$.
Indeed, the DiscB function has an imaginary part and a branch cut for $p^2 > (m_1 + m_2)^2$, 
while it is real for $p^2 <  (m_1 + m_2)^2$.
This confirms the argument above \eref{ZZZ_FzzzEFT} that  the $ZZZ$ vertex should arise from loop diagrams in the EFT where $Z$ and $h$ can simultaneously go on-shell and that the CP-violating dimension-12 operator should be present in the EFT matched at tree level to the C2HDM. 
  
 In \aref{eft}, using the functional integral methods \cite{Henning:2016lyp,Zhang:2016pja},  we sketch how to systematically derive the tree-level EFT Lagrangian for the C2HDM in the manifestly $SU(3) \times SU(2) \times U(1)$ gauge invariant language up to an arbitrary order in $1/m_H$ expansion.  
We follow that procedure and find that  the leading CP-violating operator in the bosonic sector indeed occurs at $\cO(m_H^{-8})$. 
The operator in question is identified as  
\beq
\label{eq:ZZZ_dim12}
\cL_{\rm SMEFT} \supset 
 - {Z_5^* Z_6^2 \over 2 m_H^8} \left [ D^2\left(H^\dagger X_0 \right) H \right ]^2  + \hc, 
\eeq 
where  $X_0 \equiv H^\dagger H - v^2/2$. 
Expanding the Higgs doublet around its VEV, the operator in \eref{ZZZ_dim12} yields (among others)  the $Z^3 h$ interaction term:   
{  
\beq
\label{eq:ZZZ_zzzh}
\cL_{\rm SMEFT} \supset  
\im (Z_5^* Z_6^2)  \left ( g \over c_W \right )^3  { v^7 \over 8  m_H^8}  \partial_\nu h Z^\nu  Z_\mu Z^\mu, 
\eeq }
which is P-even and C-odd (thus CP-odd). 
Hence, the dimension-12 operator in \eref{ZZZ_dim12} {  leads to CP violation when 
$\im (Z_5^* Z_6^2)  \neq 0$.
An equivalent way to derive the effective interaction in \eref{ZZZ_zzzh} from the C2HDM is to consider  a tree-level exchange of the heavy Higgs scalars between the $ZZ$ and $Zh$ vertices.  
}

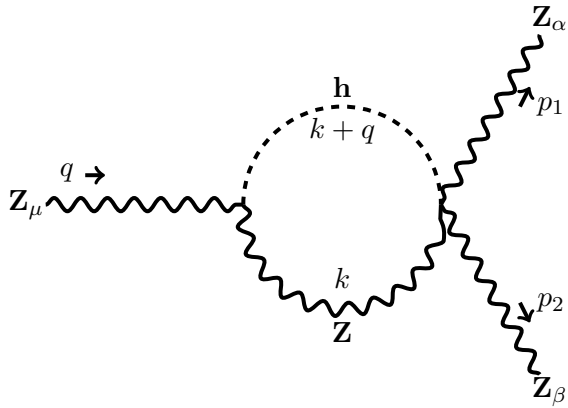
\begin{figure}
\bc
\begin{tikzpicture}[line width=1.5 pt, scale=1.3]
	\draw[vector] (180:4)--(-2,0); 
	\draw[vector] (60:2)--(0,0);
	\draw[vector] (300:2)--(0,0);
	\node at (180:4.2) {$\bf Z_\mu$};
	\node at (175:3.8) {$q$};
         \draw[->] (-3.6,0.3) -- (-3.4,0.3);
	\node at (60:2.2) {$\bf Z_\alpha$};
	\node at (42:1.5) {$p_1$};
	\draw[->] (0.8,1.0) -- (0.9,1.2);
	\node at (300:2.2) {$\bf Z_\beta$};	
	\node at (318:1.5) {$p_2$};
        \draw[->] (0.8,-1.0) -- (0.9,-1.2);
        \draw[scalarnoarrow] (-2,0) arc (180:0:1);
        \draw[vector] (-2,0) arc (180:360:1);
        \node at (-1,1.2) {$\bf h$};
        \node at (-1,0.75) {$k+q$};
        \node at (-1,-1.3) {$\bf Z$};
          \node at (-1,-0.75) {$k$};
\end{tikzpicture}
 \ec 
 \caption{
 \label{fig:zzzeft}
 A one-loop diagram contributing to the $Z^3$ vertex in the EFT.
 There are two other diagrams corresponding to permutations of the external legs.}
 \end{figure}

In the presence of the CP-violating interaction in \eref{ZZZ_zzzh}, one indeed finds 1-loop contributions to the $Z^3$ vertex,  see the diagrams in \fref{zzzeft}. 
Working in the unitary gauge we find 
{ 
\bea
 \label{eq:ZZZ_GammaEFT} 
 &  i \Gamma_{\mu \alpha \beta}  =  &
 i \im (Z_5^* Z_6^2) \left ( g \over c_W \right )^3 {v^6 \over 2 m_H^8} 
 \int {d^4 k \over (2\pi)^4} {1 \over k^2 - m_Z^2} \left \{ 
\right . \nnl   &+ &   \left.
{m_Z^2 \left [  \eta_{\mu \alpha} (k+ q)_\beta +  \eta_{\mu \beta} (k+ q)_\alpha \right ]
- k_\mu \left [ k_\alpha (k+ q)_\beta + k_\beta (k+ q)_\alpha \right ]   \over (k + q)^2 - m_h^2}
\right . \nnl   &+ &   \left.
{m_Z^2  \left [  \eta_{\mu \alpha} (k-  p_2)_\beta +  \eta_{\mu \beta} (k- p_2)_\alpha)  \right ]
- k_\beta  \left [  k_\alpha (k- p_2)_\mu  + k_\mu (k- p_2)_\alpha +  \eta_{\mu \alpha} (k^2 - kp_2)  \right ]  \over (k - p_2)^2 - m_h^2}
\right . \nnl   &+ &   \left.
{m_Z^2  \left [  \eta_{\mu \alpha} (k-  p_1)_\beta +  \eta_{\mu \beta} (k- p_1)_\alpha)  \right ]
- k_\alpha  \left [  k_\beta (k- p_1)_\mu  + k_\mu (k- p_1)_\beta +  \eta_{\mu \beta} (k^2 - kp_1)  \right ]  \over (k - p_1)^2 - m_h^2}
 \right \}  
 \nnl &+&  ({\rm ILS}). 
\eea }
Evaluating  the integral in \texttt{Package-X}~\cite{Patel:2015tea} and extracting $f_4^Z$,  we exactly recover the result in \eref{ZZZ_FzzzEFT}. 
This confirms that the dimension-12 operator in \eref{ZZZ_dim12} fully accounts for the leading $1/m_H$ behavior of the $ZZZ$ vertex in the C2HDM at one loop in the decoupling limit.  

It may be surprising that the $Z^3$ vertex in the EFT arises only at the dimension-12 level. 
After all, there are lower-dimensional CP-violating operators that lead to $Z^3$ interactions.
It is well known that the $Z^3$ vertex  cannot be generated by dimension-6 operators, 
however it does arise from the dimension-8 operator 
\beq 
\label{eq:ZZZ_dim8}
\cl_{D=8}  \supset 
{i c_{8} \over \Lambda^4}  B_{\mu \nu} B^{\mu \rho}  H^\dagger \{ D^\nu, D_\rho \} H,   
\eeq 
and other similar operators with $B_{\mu \nu} \to W_{\mu \nu}^i$ \cite{Degrande:2013kka}. 
These operators lead  to the contact $Z^3$ interaction in \eref{lZZZ}, and thus directly contribute to the  $f_4^Z$ form factor without  going through a loop diagram. 
However, one can prove that the dimension-8 operators like the one in \eref{ZZZ_dim8} cannot be generated from the C2HDM at one loop. 
The underlying reason is that  in the C2HDM all new CP-violating effects are proportional to the Jarlskog-type invariant \cite{Lavoura:1994fv}: 
\beq
\label{eq:CHDM_jarlskog}
J_{\rm CP} \equiv  {(m_{h_3}^2 - m_{h_2}^2) (m_{h_3}^2 - m_{h_1}^2) (m_{h_2}^2 - m_{h_1}^2)
\over m_{h_3}^2  m_{h_2}^2 m_{h_1}^2 } x_1 x_2 x_3 . 
\eeq 
 In the decoupling limit, \eref{ZZZ_jarlskogdecoupling} shows that this invariant is proportional to the Higgs potential couplings $Z_i$ in the 3rd power. 
The rest follows from power counting using the Planck constant $\hbar$ as a proxy \cite{Manohar:1983md,Luty:1997fk,Cohen:1997rt,Giudice:2007fh,Contino:2016jqw}. 
Reinstating the Planck constant $\hbar$ in the path integral, $\int D \phi e^{i \int d^4 x \cL/\hbar}$, 
the Lagrangian  should carry the dimension $[\cL] = \hbar^1$. 
One can assign the power $\hbar^{1/2}$ to each propagating field, the power $\hbar^{1-n/2}$ to the coupling multiplying the term with $n$ fields in the Lagrangian, and the power $\hbar^1$ for each loop factor ${1 \over 16 \pi^2}$.  
In this scheme, the electroweak couplings carry the power $[g] =  \hbar^{-1/2}$, while for the quartic Higgs couplings $[Z_i] = \hbar^{-1}$. 
It follows that the CP violating invariant is proportional to $J_{\rm CP} \sim \hbar^{-3}$. 
On the other hand, the Wilson coefficient of the dimension-8 operator  in \eref{ZZZ_dim8} should have $[c_8] = \hbar^{-1}$.  
If that operator arises at $l$ loops in the C2HDM then $c_8 \sim {g^2 J_{\rm CP} \over (16 \pi^2)^l}$ and the $\hbar$ power counting fixes $l = 3$: the dimension-8 operator in  \eref{ZZZ_dim8} cannot appear before the 3 loop level in the matching of  the SMEFT to the C2HDM.   
The same power counting shows that the dimension-12 operator in \eref{ZZZ_dim8} is allowed at tree level.

\section{Conclusions}
\label{sec:conclusions}

In this paper we have studied the CP violating triple-Z vertex in the C2HDM and in its effective description  within the SMEFT framework.
From the point of view of the high-energy theory, the $ZZZ$ vertex probes one of the two independent Jarlskog-type invariants in the extended Higgs sector. 
The leading contributions arise from triangle one-loop diagrams with both SM particles and the new Higgs scalars. 
We obtained the CP violating form factor $f_4^Z$ in a general $R_\xi$ gauge, thus demonstrating the gauge invariance of the result and  reassessing previous calculations in the literature.  
Starting from the (complicated) full result, we extracted an analytic approximation valid in the decoupling limit when the mass scale $m_H$ of the heavy scalars is much larger than $m_Z$ and the momentum flowing through the vertex. 
Given that approximation, we were able to identify the operators and diagrams responsible for the generation of the $ZZZ$ vertex in the low-energy effective theory where the heavy scalars are integrated out. 
Even though the $ZZZ$ vertex can in principle be generated by dimension-8 operators in the SMEFT, 
such contributions are absent in the effective theory matched to the C2HDM at one loop. 
This fact may be surprising at first, but it follows from simple power counting, given the dependence of the Jarlskog invariants on the masses and couplings of the C2HDM.   
Instead, we found that the CP violating $ZZZ$ vertex appears in the effective theory only at the level of dimension-12 operators.    
In practice, this means the CP violating effects in diboson production will be extremely suppressed  (by a loop factor multiplied by $(v/m_H)^8$) if the mass scale of the heavy Higgs partners is well above the weak scale.

\acknowledgments
We are grateful to Celine Degrande and to Alexander Savin for discussions.
This work was funded in part by a bilateral agreement,
sponsored in Portugal by FCT -
\textit{Funda\c{c}\~{a}o para a Ci\^{e}ncia e a Tecnologia},
under program HIGGS E SABOR FCT " 2015/2016 - PROC 441.00 FRANCA,
and sponsored in France by the {\em Partenariats Hubert Curien} programme PESSOA  under project  33733UH. 
A.F is partially supported by the ERC Advanced Grant Higgs@LHC and by the European Union's Horizon 2020 research and innovation programme under the Marie Sklodowska-Curie grant agreements No 690575 and  No 674896.
J.C.R.~and J.P.S.~are supported in part by the Portuguese FCT under contract No. UID/FIS/00777/2013.

\appendix 

\section{Effective $ZZZ$ vertex via method of regions}
\label{app:mr}

In this appendix we discuss how to isolate the leading contribution to the CP-violating $ZZZ$ form factor 
$f_4^Z(q^2)$ in the C2HDM in the limit where the extra scalars are much heavier than the Higgs boson.  
To this end we will utilize the method of regions~\cite{Beneke:1997zp}.
A loop integral with two disparate mass scales $m_{\rm light} \ll m_{\rm heavy}$ can be calculated by A) expanding the integrand for {\em soft} momenta $k \sim m_{\rm light}$ and performing the integral, 
B) expanding the integrand for {\em hard} momenta $k \sim m_{\rm heavy}$ and performing the integral,
and adding the contributions A) and B).  
Note that the separate soft and hard contributions may be UV or IR divergent. 
If that is the case, the integrals have to be regulated,  with the regulator dependence canceling out when the full result is finite.  
In the following we will implicitly use the dimensional regularization which is convenient because the  EFT expansion is not  complicated by the presence of  massive regulators.  

We apply this method to the $ZZZ$ vertex in the C2HDM, whose integral representation is given in  
\eref{ZZZ_GammaFinal}. 
For presentation purposes, in this appendix we work in the unphysical limit $m_1 = m_Z$. 
The reason is that in this limit the $ZZZ$ vertex simplifies considerably, 
as  the first four lines in \eref{ZZZ_GammaFinal} cancel against each other: 
\bea 
 \label{eq:ZZZ_GammaApp} 
i \Gamma_{\mu \alpha \beta}  & \to &
- i  m_Z^2 \left ( g \over c_W \right ) ^3 (x_1 x_2 x_3)  \int {d^4 k \over (2\pi)^4}  \eps_{ijk} \left \{ 
 {k_\beta \eta_{\mu \alpha}  \over  ((k-p_1)^2 - m_Z^2 ) ( (k+p_2)^2 - m_j^2) (k^2 - m_k^2)} 
\right . \nnl   &-&   \left.
 {k_\alpha  \eta_{\mu \beta}  \over ( (k-p_1)^2 - m_j^2) ((k+p_2)^2 -  m_Z^2 ) (k^2 - m_k^2) }
 \right \}   + ({\rm ILS }) \equiv i  \hat \Gamma_{\mu \alpha \beta} . 
\eea 
Taking that simplified  limit  allows us to  illustrate the gist of the argument. 
The discussion for the  general case $m_1 = m_h$ is completely analogous, but much more tedious and paper-consuming. 

We can rewrite \eref{ZZZ_GammaApp} as 
\beq
i \hat \Gamma_{\mu \alpha \beta}  = 
-   m_Z^2 (m_3^2 - m_2^2) \left ( g \over c_W \right ) ^3 (x_1 x_2 x_3)  \left [ 
 \eta_{\mu \alpha} (I^{LH}_\beta  +  I^{HH}_\beta ) 
 +  \eta_{\mu \beta} (\tilde I^{LH}_\alpha  +  \tilde I^{HH}_\alpha )  \right ]   + ({\rm ILS }).  
\eeq 
Here $I^{LH}_\beta$ sums  the contributions with one heavy scalar in the loop: 
\bea 
I^{LH}_\beta &=  & \int {d^4 k \over (2\pi)^4}   {i k_\beta \over ((k-p_1)^2 - m_Z^2 )}
\nnl & \times & 
 \left [
{1 \over  (k^2 - m_Z^2)((k+p_2)^2 - m_2^2)((k+p_2)^2 - m_3^2)  }
- {1 \over ((k+p_2)^2 - m_Z^2)(k^2 - m_2^2) (k^2 - m_3^2)} \right ],  
\nnl 
\eea 
while $I^{HH}_\beta$ sums  the contributions with two heavy scalars in the loop: 
 \beq 
I^{HH}_\beta = \int {d^4 k \over (2\pi)^4}   {i k_\beta (2 k p_2 + m_Z^2) \over ((k-p_1)^2 - m_Z^2 )}
{1 \over  (k^2 - m_2^2)(k^2 - m_3^2)((k+p_2)^2 - m_2^2)((k+p_2)^2 - m_3^2)  }. 
\eeq   
 $\tilde I^{XH}_\alpha$ is the same as  $I^{XH}_\alpha$ with $p_2 \leftrightarrow p_1$.  
 Note that $I^{XH}$ have dimensions $[{\rm mass}]^{-4}$. 

Let us apply the method of regions to  $I^{XH}_\beta$. We assume $m_2 \sim m_3 \sim m_H \gg m_Z$. 
Starting with $I^{HH}_\beta$, the soft limit is strongly suppressed by the heavy scalar mass, 
$I^{HH, \rm soft}_\beta \sim O(m_H^{-8})$. 
In the hard limit the suppression is less severe: 
$I^{HH, \rm hard}_\beta 
\approx  \cO(m_H^{-4}) p_{2,\beta} +\cO(m_H^{-6}) p_{1,\beta}$. 
Contributions to $f_4^Z$ arise only from the second term, thus they are $\cO(m_H^{-6})$.
Turning to $I^{LH}_\beta$, its hard part scales in the same way as $I^{HH, \rm hard}$. 
However, the soft part is only suppressed by $m_H^{-4}$: 
\beq
I^{LH, \rm soft}_\beta \approx  {i \over m_H^4} \int {d^4 k \over (2\pi)^4}   {k_\beta \over ((k-p_1)^2 - m_Z^2 )}
 \left [
{1 \over  (k^2 - m_Z^2)  }
- {1 \over ((k+p_2)^2 - m_Z^2)}  \right ],  
\eeq 
and therefore provides the leading contribution to $f_4^Z(q^2)$. 
The integral can be evaluated using \texttt{Package-X}~\cite{Patel:2015tea}:  
\beq
I^{LH, \rm soft}_\beta \approx {\sqrt 3 \pi + 3 {\rm DiscB}[q^2, m_Z, m_Z] \over 96 \pi^2 m_H^4 }, 
\eeq 
where the function ${\rm DiscB}$ is defined in \eref{ZZZ_discb2}. 
All in all we find 
\beq
\label{eq:f4Zmhmz} 
e f_4^Z(q^2) = 
-  {m_Z^2 (m_3^2 - m_2^2) \over m_H^4} \left ( g \over c_W \right ) ^3 (x_1 x_2 x_3)  {\sqrt 3 \pi + 3 {\rm DiscB}[q^2, m_Z, m_Z] \over 96 \pi^2 } +  \cO(m_H^{-6}).   
\eeq 
For arbitrary $m_h$ the calculation is completely analogous.  
In particular, the leading contribution to 
$f_4^Z$ is still $\cO(m_H^{-4})$ and corresponds to the soft region of the diagrams with a single heavy scalar in the loop.  
The (much more lengthy) result in the general case is displayed   in \eref{ZZZ_FzzzEFT}. 
With a bit of algebra one can show that \eref{f4Zmhmz} indeed corresponds to the $m_h \to m_Z$ limit of the general result.  

Taking into account $x_1 x_2 x_3 \sim \cO(m_H^{-4})$, the form factor is dramatically suppressed, $f_4^Z \sim \cO(m_H^{-8})$.  
This tells us that in the low-energy EFT below the scale $m_H$ the form factor must arise from a dimension-12 operator. 
Since the leading contribution to $f_4^Z$  arises from the soft region of the integral in \eref{ZZZ_GammaApp}, 
in the EFT it will be reproduced by a loop diagram with a single insertion of the dimension-12 operator. 
The same conclusion can be reached by observing that   $f_4^Z$
has a branch cut corresponding to the light scalar and $Z$ boson in the loop simultaneously going on-shell.  
The responsible operator, the  CP-violating vertex,  and the loop diagram were identified in \sref{eft2hdm}. 
The hard part of \eref{ZZZ_GammaApp} corresponds to tree-level  contributions of contact interactions  in the EFT, but that is suppressed by an additional  factor of $1/m_H^2$ and thus enters only at the level of dimension-14 operators.

\section{$ZZZ$ vertex from CP-violating EFT operators}
\label{app:eft}

In this appendix we identify the leading CP-violating operators contributing to the $ZZZ$ vertex in the low-energy effective theory of the C2HDM after integrating out the heavy Higgs scalars. 
An efficient way to proceed is to use functional methods while keeping the electroweak $SU(2)\times U(1)$ symmetry manifest. 
In this approach,  the effective Lagrangian at tree level  is given by 
$\cL_{\rm EFT}(H_1) = \cL_{\rm C2HDM}(H_1,H_2^c(H_1))$,  
where $H_2^c$ is the solution of its classical equation of motion in the C2HDM with $H_1$ treated as background field. 
The result does not depend on which 2HDM basis is used as the starting point, however the calculation is simplest in the Higgs basis where one avoids the complications of VEV and couplings redefinitions and $Z h$ kinetic mixing. 

Solving the equation of motion and deriving $\cL_{\rm EFT}$  can be readily performed perturbatively in the $1/Y_2$ expansion. 
As explained in \sref{eft2hdm}, power counting arguments show that the relevant CP-violating operators arise only at the level at $\cO(Y_2^{-4})$, corresponding to  dimension-12 operators. 
Deriving the complete effective Lagrangian up to dimension 12 would be quite a task.  
For the present purpose, we focus only on its small fragment containing  purely bosonic CP-violating interactions. 
We know that these have to be proportional to the Jarlskog-like invariant $\im(Z_5^* Z_6^2)$.
Therefore we will only trace the terms in  $\cL_{\rm EFT}$ that contain $Z_5$ or $Z_6$, and ignore everything else. 
We thus consider the C2HDM  Lagrangian 
\bea 
\label{eq:lc2hdm}
\cL_{\rm C2HDM} &=& |D_\mu H_1|^2  -  Y_1 |H_1|^2  -  \frac{Z_1}{2} |H_1|^4
\\  & + &  |D_\mu H_2|^2 -  Y_2 |H_2|^2   -  \left [  (Y_3 + Z_6 |H_1|^2) H_1^\dagger H_2 + \hc \right ]
-   \left [ \frac{Z_5}{2} (H_1^\dagger H_2)^2 + \textrm{h.c.} \right ] + \dots , 
\nonumber 
\eea
where the dots stand for other terms in the Higgs potential, gauge kinetic terms, and all fermionic terms, which are not relevant for the present discussion. 
We work in the Higgs basis where $Y_3 = - Z_6 v^2/2$,  $\langle H_1 \rangle = v/\sqrt{2}$, $\langle H_2 \rangle = 0$.  
Then $Y_3 + Z_6 |H_1|^2 = Z_6 X_0$,  where we defined $X_0 \equiv |H_1|^2 - v^2/2$.   
The equation of motion for $H_2$ takes the form 
\beq
Y_2 H_2 + D^2 H_2 + Z_6^* X_0 H_1 + Z_5^* (H_2^\dagger H_1) H_1 + \dots  = 0. 
\eeq
We search for a perturbative solution in the form $H_2^c = \sum_{n=1}^\infty Y_2^{-n} H_2^{(n)}$. 
This leads to the recursive  system  of  equations: 
\bea
\label{eq:eomrec}
H_2^{(1)}  & = & - Z_6^* X_0 H_1 + \dots, 
\nnl 
H_2^{(n+1)}  & = &  - D^2 H_2^{(n)} - Z_5^* (H_2^{(n)}{}^\dagger H_1) H_1  +  \dots, 
\eea 
which determines $H_2^c$ (or at least its part depending on $Z_5$ and $Z_6$) to an arbitrary order $n$.
We will only need the explicit solution up to $n=4$: 
\bea
\label{eq:eomsol}
H_2^{(2)}  & =  & Z_6^* D^2 (X_0 H_1)  + Z_5^* Z_6 X_0 |H_1|^2 H_1 + \dots, 
\\   
H_2^{(4)}  & =  & D^4 H_2^{(2)} + Z_5^* D^2 [ (H_2^{(2)}{}^\dagger H_1) H_1 ] 
+ Z_5^* ( D^2 H_2^{(2)}{}^\dagger H_1) H_1 + |Z_5|^2 |H_1|^2  (H_1^\dagger   H_2^{(2)}) H_1 + \dots . 
\nonumber
\eea 
 Plugging that solution back into \eref{lc2hdm} one gets the EFT Lagrangian in $1/Y_2$ expansion: 
$\cL_{\rm EFT} = \cL_{\rm SM} + \sum_{n = 1}^\infty Y_2^{-n}\cL^{(2n +4)}$, 
where each term contains local operators composed of $H_1$ and its (covariant) derivatives.  
In the low-energy theory  $H_2$ is integrated out and $H_1$ remains as the only doublet scalar, so in the following we relabel $H_1 \to H$. 
It is now a trivial if tedious exercise to determine the EFT operators  $\cL^{(2n +4)}$ at each given order. 
For example, this procedure yields $\cL^{(6)} \supset  |Z_6|^2 X_0^2 |H|^2$ which shifts the triple Higgs boson coupling away from the SM prediction,  or  $\cL^{(8)} \supset  |Z_6 D_\mu (X_0 H)|^2 $ which renormalizes the Higgs boson kinetic terms, thus uniformly shifting all the Higgs boson couplings. 
At $n=8$ we also  encounter a term proportional to $Z_5^* Z_6^2$,    
$\cL^{(8)} \supset  -{Z_5^* Z_6^2 X_0^2  |H|^4 \over 2} + \hc$,
but it is  CP-conserving and only yields interactions proportional to $\re Z_5^* Z_6^2$.
At $n =10$ we find 
$\cL^{(10)} \supset  -{Z_5^* Z_6^2 D_\mu(H^\dagger X_0) D_\mu (X_0 |H|^2 H) \over 2} + \hc$, 
but the only resulting interactions proportional to $\im Z_5^* Z_6^2$ are of the form $\sim h^m \partial_\mu Z^\mu$ with $m \geq 3$, which is not interesting for our purpose.  
The first time we encounter a genuine CP-violating interaction   proportional   to $\im Z_5^* Z_6^2$
is in  $\cL^{(12)}$. 
At that order the effective Lagrangian can be written as 
\bea
 \cL^{(12)}  &=&   - {1\over 2} H_2^{(2)}{}^\dagger D^2 H_2^{(2)}  
-  H_2^{(1)}{}^\dagger D^2 H_2^{(3)} 
-  H_2^{(1)}{}^\dagger H_2^{(4)}   - H_2^{(2)}{}^\dagger H_2^{(3)}  
\nnl 
&-  & Z_6 X_0 H^\dagger H_2^{(4)}
-  \frac{Z_5}{2} (H^\dagger H_2^{(2)})^2 
-  Z_5   (H^\dagger H_2^{(1)}) (H^\dagger H_2^{(3)}) + \hc 
\nnl  & = & 
  {1\over 2} H_2^{(2)}{}^\dagger D^2 H_2^{(2)}  
  -  Z_6 X_0 H^\dagger H_2^{(4)}
  + \frac{Z_5}{2} (H^\dagger H_2^{(2)})^2    + \hc + \dots 
\eea 
To derive the second equality we used the recursion in \eref{eomrec}. 
Plugging in the solution in \eref{eomsol} and integrating by parts we find that $\cL^{(12)}$ contains the following terms  proportional to $Z_5^* Z_6^2$: 
\beq
 \cL^{(12)}   \supset - Z_5^* Z_6^2 \left [ D^2(H^\dagger X_0) D^2 (X_0 |H|^2 H)
 + {1 \over 2} \left ( D^2(H^\dagger X_0) H \right )^2 \right ] + \hc . 
\eeq 
Ignoring again interactions proportional to $\partial_\mu Z^\mu$, 
only the second term in the bracket leads to non-trivial CP-violating interactions: 
\bea
 \cL^{(12)}  &  \supset &  \im (Z_5^* Z_6^2)  {g v^6 \over 2  c_W}   Z^\nu   \partial_\nu h \, \Box h 
 + \cO(Z h^3) 
 \nnl & \to & 
  \im (Z_5^* Z_6^2)  {g v^5 \over 2  c_W}  \partial_\nu h Z^\nu \left ( m_Z^2 Z_\mu Z^\mu + 2 m_W^2 W_\mu^+ W^{\mu \, -} \right )  . 
\eea 
In the last step we used the classical equation of motion for the Higgs boson field.  
These are the leading CP-violating interactions in the bosonic sector of the low-energy effective theory of C2HDM.  
At one loop in the EFT, the interaction term $\sim \partial_\nu h Z^\nu  Z_\mu Z_\mu$ generates the CP-violating $ZZZ$ vertex via the Feynman diagram in \fref{zzzeft}.

\bibliographystyle{JHEP} 
\bibliography{BFFRS}

\providecommand{\href}[2]{#2}\begingroup\raggedright\begin{thebibliography}{10}

\bibitem{Aad:2012tfa}
{\bf ATLAS} Collaboration, G.~Aad et~al., {\it {Observation of a new particle
  in the search for the Standard Model Higgs boson with the ATLAS detector at
  the LHC}},  {\em Phys.Lett.} {\bf B716} (2012) 1--29,
  [\href{http://arxiv.org/abs/1207.7214}{{\tt arXiv:1207.7214}}].

\bibitem{Chatrchyan:2012xdj}
{\bf CMS} Collaboration, S.~Chatrchyan et~al., {\it {Observation of a new boson
  at a mass of 125 GeV with the CMS experiment at the LHC}},  {\em Phys. Lett.}
  {\bf B716} (2012) 30--61, [\href{http://arxiv.org/abs/1207.7235}{{\tt
  arXiv:1207.7235}}].

\bibitem{Higgs:1964ia}
P.~W. Higgs, {\it {Broken symmetries, massless particles and gauge fields}},
  {\em Phys.Lett.} {\bf 12} (1964) 132--133.

\bibitem{Higgs:1964pj}
P.~W. Higgs, {\it {Broken Symmetries and the Masses of Gauge Bosons}},  {\em
  Phys.Rev.Lett.} {\bf 13} (1964) 508--509.

\bibitem{Englert:1964et}
F.~Englert and R.~Brout, {\it {Broken Symmetry and the Mass of Gauge Vector
  Mesons}},  {\em Phys.Rev.Lett.} {\bf 13} (1964) 321--323.

\bibitem{Guralnik:1964eu}
G.~Guralnik, C.~Hagen, and T.~Kibble, {\it {Global Conservation Laws and
  Massless Particles}},  {\em Phys.Rev.Lett.} {\bf 13} (1964) 585--587.

\bibitem{Glashow:1961tr}
S.~L. Glashow, {\it {Partial Symmetries of Weak Interactions}},  {\em Nucl.
  Phys.} {\bf 22} (1961) 579--588.

\bibitem{Weinberg:1967tq}
S.~Weinberg, {\it {A Model of Leptons}},  {\em Phys.Rev.Lett.} {\bf 19} (1967)
  1264--1266.

\bibitem{Gunion:1989we}
J.~F. Gunion, H.~E. Haber, G.~L. Kane, and S.~Dawson, {\it {The Higgs Hunter's
  Guide}},  {\em Front. Phys.} {\bf 80} (2000) 1--404.

\bibitem{Branco:2011iw}
G.~C. Branco, P.~M. Ferreira, L.~Lavoura, M.~N. Rebelo, M.~Sher, and J.~P.
  Silva, {\it {Theory and phenomenology of two-Higgs-doublet models}},  {\em
  Phys. Rept.} {\bf 516} (2012) 1--102,
  [\href{http://arxiv.org/abs/1106.0034}{{\tt arXiv:1106.0034}}].

\bibitem{Ginzburg:2002wt}
I.~F. Ginzburg, M.~Krawczyk, and P.~Osland, {\it {Two Higgs doublet models with
  CP violation}},  in {\em {Linear colliders. Proceedings, International
  Workshop on physics and experiments with future electron-positron linear
  colliders, LCWS 2002, Seogwipo, Jeju Island, Korea, August 26-30, 2002}},
  pp.~703--706, 2002.
\newblock \href{http://arxiv.org/abs/hep-ph/0211371}{{\tt hep-ph/0211371}}.

\bibitem{Khater:2003wq}
W.~Khater and P.~Osland, {\it {CP violation in top quark production at the LHC
  and two Higgs doublet models}},  {\em Nucl. Phys.} {\bf B661} (2003)
  209--234, [\href{http://arxiv.org/abs/hep-ph/0302004}{{\tt hep-ph/0302004}}].

\bibitem{ElKaffas:2007rq}
A.~W. El~Kaffas, P.~Osland, and O.~M. Ogreid, {\it {CP violation, stability and
  unitarity of the two Higgs doublet model}},  {\em Nonlin. Phenom. Complex
  Syst.} {\bf 10} (2007) 347--357,
  [\href{http://arxiv.org/abs/hep-ph/0702097}{{\tt hep-ph/0702097}}].

\bibitem{ElKaffas:2006gdt}
A.~W. El~Kaffas, W.~Khater, O.~M. Ogreid, and P.~Osland, {\it {Consistency of
  the two Higgs doublet model and CP violation in top production at the LHC}},
  {\em Nucl. Phys.} {\bf B775} (2007) 45--77,
  [\href{http://arxiv.org/abs/hep-ph/0605142}{{\tt hep-ph/0605142}}].

\bibitem{Grzadkowski:2009iz}
B.~Grzadkowski and P.~Osland, {\it {Tempered Two-Higgs-Doublet Model}},  {\em
  Phys. Rev.} {\bf D82} (2010) 125026,
  [\href{http://arxiv.org/abs/0910.4068}{{\tt arXiv:0910.4068}}].

\bibitem{Arhrib:2010ju}
A.~Arhrib, E.~Christova, H.~Eberl, and E.~Ginina, {\it {CP violation in charged
  Higgs production and decays in the Complex Two Higgs Doublet Model}},  {\em
  JHEP} {\bf 04} (2011) 089, [\href{http://arxiv.org/abs/1011.6560}{{\tt
  arXiv:1011.6560}}].

\bibitem{Barroso:2012wz}
A.~Barroso, P.~M. Ferreira, R.~Santos, and J.~P. Silva, {\it {Probing the
  scalar-pseudoscalar mixing in the 125 GeV Higgs particle with current data}},
   {\em Phys. Rev.} {\bf D86} (2012) 015022,
  [\href{http://arxiv.org/abs/1205.4247}{{\tt arXiv:1205.4247}}].

\bibitem{Fontes:2014xva}
D.~Fontes, J.~C. Rom\~{a}o, and J.~P. Silva, {\it {$h \rightarrow Z \gamma$ in
  the complex two Higgs doublet model}},  {\em JHEP} {\bf 12} (2014) 043,
  [\href{http://arxiv.org/abs/1408.2534}{{\tt arXiv:1408.2534}}].

\bibitem{Gorbahn:2015gxa}
M.~Gorbahn, J.~M. No, and V.~Sanz, {\it {Benchmarks for Higgs Effective Theory:
  Extended Higgs Sectors}},  {\em JHEP} {\bf 10} (2015) 036,
  [\href{http://arxiv.org/abs/1502.07352}{{\tt arXiv:1502.07352}}].

\bibitem{Brehmer:2015rna}
J.~Brehmer, A.~Freitas, D.~L\'{o}pez-Val, and T.~Plehn, {\it {Pushing Higgs
  Effective Theory to its Limits}},  {\em Phys. Rev.} {\bf D93} (2016), no.~7
  075014, [\href{http://arxiv.org/abs/1510.03443}{{\tt arXiv:1510.03443}}].

\bibitem{Egana-Ugrinovic:2015vgy}
D.~Egana-Ugrinovic and S.~Thomas, {\it {Effective Theory of Higgs Sector Vacuum
  States}},  \href{http://arxiv.org/abs/1512.00144}{{\tt arXiv:1512.00144}}.

\bibitem{Freitas:2016iwx}
A.~Freitas, D.~L\'{o}pez-Val, and T.~Plehn, {\it {When matching matters: Loop
  effects in Higgs effective theory}},  {\em Phys. Rev.} {\bf D94} (2016),
  no.~9 095007, [\href{http://arxiv.org/abs/1607.08251}{{\tt
  arXiv:1607.08251}}].

\bibitem{Belusca-Maito:2016dqe}
H.~B{\'e}lusca-Ma{\"i}to, A.~Falkowski, D.~Fontes, J.~C. Rom\~{a}o, and J.~P.
  Silva, {\it {Higgs EFT for 2HDM and beyond}},  {\em Eur. Phys. J.} {\bf C77}
  (2017), no.~3 176, [\href{http://arxiv.org/abs/1611.01112}{{\tt
  arXiv:1611.01112}}].

\bibitem{Corbett:2017ieo}
T.~Corbett, A.~Joglekar, H.-L. Li, and J.-H. Yu, {\it {Exploring Extended
  Scalar Sectors with Di-Higgs Signals: A Higgs EFT Perspective}},
  \href{http://arxiv.org/abs/1705.02551}{{\tt arXiv:1705.02551}}.

\bibitem{Karmakar:2017yek}
S.~Karmakar and S.~Rakshit, {\it {Higher dimensional operators in 2HDM}},
  \href{http://arxiv.org/abs/1707.00716}{{\tt arXiv:1707.00716}}.

\bibitem{Grzadkowski:2016lpv}
B.~Grzadkowski, O.~M. Ogreid, and P.~Osland, {\it {CP-Violation in the $ZZZ$
  and $ZWW$ vertices at $e^+e^-$ colliders in Two-Higgs-Doublet Models}},  {\em
  JHEP} {\bf 05} (2016) 025, [\href{http://arxiv.org/abs/1603.01388}{{\tt
  arXiv:1603.01388}}].

\bibitem{Lavoura:1994fv}
L.~Lavoura and J.~P. Silva, {\it {Fundamental CP violating quantities in a
  SU(2) x U(1) model with many Higgs doublets}},  {\em Phys. Rev.} {\bf D50}
  (1994) 4619--4624, [\href{http://arxiv.org/abs/hep-ph/9404276}{{\tt
  hep-ph/9404276}}].

\bibitem{Botella:1994cs}
F.~J. Botella and J.~P. Silva, {\it {Jarlskog - like invariants for theories
  with scalars and fermions}},  {\em Phys. Rev.} {\bf D51} (1995) 3870--3875,
  [\href{http://arxiv.org/abs/hep-ph/9411288}{{\tt hep-ph/9411288}}].

\bibitem{delAguila:2016zcb}
F.~del Aguila, Z.~Kunszt, and J.~Santiago, {\it {One-loop effective lagrangians
  after matching}},  {\em Eur. Phys. J.} {\bf C76} (2016), no.~5 244,
  [\href{http://arxiv.org/abs/1602.00126}{{\tt arXiv:1602.00126}}].

\bibitem{Henning:2016lyp}
B.~Henning, X.~Lu, and H.~Murayama, {\it {One-loop Matching and Running with
  Covariant Derivative Expansion}},
  \href{http://arxiv.org/abs/1604.01019}{{\tt arXiv:1604.01019}}.

\bibitem{Ellis:2016enq}
S.~A.~R. Ellis, J.~Quevillon, T.~You, and Z.~Zhang, {\it {Mixed heavy-light
  matching in the Universal One-Loop Effective Action}},  {\em Phys. Lett.}
  {\bf B762} (2016) 166--176, [\href{http://arxiv.org/abs/1604.02445}{{\tt
  arXiv:1604.02445}}].

\bibitem{Fuentes-Martin:2016uol}
J.~Fuentes-Martin, J.~Portoles, and P.~Ruiz-Femenia, {\it {Integrating out
  heavy particles with functional methods: a simplified framework}},  {\em
  JHEP} {\bf 09} (2016) 156, [\href{http://arxiv.org/abs/1607.02142}{{\tt
  arXiv:1607.02142}}].

\bibitem{Zhang:2016pja}
Z.~Zhang, {\it {Covariant diagrams for one-loop matching}},  {\em JHEP} {\bf
  05} (2017) 152, [\href{http://arxiv.org/abs/1610.00710}{{\tt
  arXiv:1610.00710}}].

\bibitem{Ellis:2017jns}
S.~A.~R. Ellis, J.~Quevillon, T.~You, and Z.~Zhang, {\it {Extending the
  Universal One-Loop Effective Action: Heavy-Light Coefficients}},  {\em JHEP}
  {\bf 08} (2017) 054, [\href{http://arxiv.org/abs/1706.07765}{{\tt
  arXiv:1706.07765}}].

\bibitem{Gounaris:2006zm}
G.~J. Gounaris and F.~M. Renard, {\it {Addendum to `Helicity conservation in
  gauge boson scattering at high energy'}},  {\em Phys. Rev.} {\bf D73} (2006)
  097301, [\href{http://arxiv.org/abs/hep-ph/0604041}{{\tt hep-ph/0604041}}].

\bibitem{Degrande:2013kka}
C.~Degrande, {\it {A basis of dimension-eight operators for anomalous neutral
  triple gauge boson interactions}},  {\em JHEP} {\bf 02} (2014) 101,
  [\href{http://arxiv.org/abs/1308.6323}{{\tt arXiv:1308.6323}}].

\bibitem{Hagiwara:1986vm}
K.~Hagiwara, R.~D. Peccei, D.~Zeppenfeld, and K.~Hikasa, {\it {Probing the Weak
  Boson Sector in $e^+ e^- \to W^+ W^-$}},  {\em Nucl. Phys.} {\bf B282} (1987)
  253--307.

\bibitem{Gounaris:1999kf}
G.~J. Gounaris, J.~Layssac, and F.~M. Renard, {\it {Signatures of the anomalous
  $Z_{\gamma}$ and $Z Z$ production at the lepton and hadron colliders}},  {\em
  Phys. Rev.} {\bf D61} (2000) 073013,
  [\href{http://arxiv.org/abs/hep-ph/9910395}{{\tt hep-ph/9910395}}].

\bibitem{Chang:1994cs}
D.~Chang, W.-Y. Keung, and P.~B. Pal, {\it {CP violation in the cubic coupling
  of neutral gauge bosons}},  {\em Phys. Rev.} {\bf D51} (1995) 1326--1331,
  [\href{http://arxiv.org/abs/hep-ph/9407294}{{\tt hep-ph/9407294}}].

\bibitem{Glashow:1976nt}
S.~L. Glashow and S.~Weinberg, {\it {Natural Conservation Laws for Neutral
  Currents}},  {\em Phys. Rev.} {\bf D15} (1977) 1958.

\bibitem{Paschos:1976ay}
E.~A. Paschos, {\it {Diagonal Neutral Currents}},  {\em Phys. Rev.} {\bf D15}
  (1977) 1966.

\bibitem{Gunion:2002zf}
J.~F. Gunion and H.~E. Haber, {\it {The CP conserving two Higgs doublet model:
  The Approach to the decoupling limit}},  {\em Phys. Rev.} {\bf D67} (2003)
  075019, [\href{http://arxiv.org/abs/hep-ph/0207010}{{\tt hep-ph/0207010}}].

\bibitem{Bernon:2015qea}
J.~Bernon, J.~F. Gunion, H.~E. Haber, Y.~Jiang, and S.~Kraml, {\it
  {Scrutinizing the alignment limit in two-Higgs-doublet models:
  $m_h=125$~GeV}},  {\em Phys. Rev.} {\bf D92} (2015), no.~7 075004,
  [\href{http://arxiv.org/abs/1507.00933}{{\tt arXiv:1507.00933}}].

\bibitem{Passarino:1978jh}
G.~Passarino and M.~J.~G. Veltman, {\it {One Loop Corrections for $e^+ e^-$
  Annihilation Into $\mu^+ \mu^-$ in the Weinberg Model}},  {\em Nucl. Phys.}
  {\bf B160} (1979) 151--207.

\bibitem{Hahn:1998yk}
T.~Hahn and M.~Perez-Victoria, {\it {Automatized one loop calculations in
  four-dimensions and D-dimensions}},  {\em Comput. Phys. Commun.} {\bf 118}
  (1999) 153--165, [\href{http://arxiv.org/abs/hep-ph/9807565}{{\tt
  hep-ph/9807565}}].

\bibitem{Mertig:1990an}
R.~Mertig, M.~Bohm, and A.~Denner, {\it {FEYN CALC: Computer algebraic
  calculation of Feynman amplitudes}},  {\em Comput. Phys. Commun.} {\bf 64}
  (1991) 345--359.

\bibitem{Patel:2015tea}
H.~H. Patel, {\it {Package-X: A Mathematica package for the analytic
  calculation of one-loop integrals}},  {\em Comput. Phys. Commun.} {\bf 197}
  (2015) 276--290, [\href{http://arxiv.org/abs/1503.01469}{{\tt
  arXiv:1503.01469}}].

\bibitem{WebPageC2HDM}
D.~Fontes, M.~M{\"u}hlleitner, J.~Rom\~{a}o, R.~Santos, J.~P. Silva, and
  J.~Wittbrodt. \url{http://porthos.tecnico.ulisboa.pt/arXiv/C2HDM/}, 2017.

\bibitem{Romao:2012pq}
J.~C. Rom\~{a}o and J.~P. Silva, {\it {A resource for signs and Feynman
  diagrams of the Standard Model}},  {\em Int. J. Mod. Phys.} {\bf A27} (2012)
  1230025, [\href{http://arxiv.org/abs/1209.6213}{{\tt arXiv:1209.6213}}].

\bibitem{Aaboud:2017rwm}
{\bf ATLAS} Collaboration, M.~Aaboud et~al., {\it {$ZZ \to
  \ell^{+}\ell^{-}\ell^{\prime +}\ell^{\prime -}$ cross-section measurements
  and search for anomalous triple gauge couplings in 13 TeV $pp$ collisions
  with the ATLAS detector}},  \href{http://arxiv.org/abs/1709.07703}{{\tt
  arXiv:1709.07703}}.

\bibitem{Sirunyan:2017zjc}
{\bf CMS} Collaboration, A.~M. Sirunyan et~al., {\it {Measurements of the
  pp$\to$ZZ production cross section and the Z$\to 4\ell$ branching fraction,
  and constraints on anomalous triple gauge couplings at $\sqrt{s} =$ 13 TeV}},
   \href{http://arxiv.org/abs/1709.08601}{{\tt arXiv:1709.08601}}.

\bibitem{Baron:2013eja}
{\bf ACME} Collaboration, J.~Baron et~al., {\it {Order of Magnitude Smaller
  Limit on the Electric Dipole Moment of the Electron}},  {\em Science} {\bf
  343} (2014) 269--272, [\href{http://arxiv.org/abs/1310.7534}{{\tt
  arXiv:1310.7534}}].

\bibitem{Perez:1995dc}
M.~A. Perez, J.~J. Toscano, and J.~Wudka, {\it {Two photon processes and
  effective Lagrangians with an extended scalar sector}},  {\em Phys. Rev.}
  {\bf D52} (1995) 494--504, [\href{http://arxiv.org/abs/hep-ph/9506457}{{\tt
  hep-ph/9506457}}].

\bibitem{Henning:2014wua}
B.~Henning, X.~Lu, and H.~Murayama, {\it {How to use the Standard Model
  effective field theory}},  {\em JHEP} {\bf 01} (2016) 023,
  [\href{http://arxiv.org/abs/1412.1837}{{\tt arXiv:1412.1837}}].

\bibitem{Beneke:1997zp}
M.~Beneke and V.~A. Smirnov, {\it {Asymptotic expansion of Feynman integrals
  near threshold}},  {\em Nucl. Phys.} {\bf B522} (1998) 321--344,
  [\href{http://arxiv.org/abs/hep-ph/9711391}{{\tt hep-ph/9711391}}].

\bibitem{Manohar:1983md}
A.~Manohar and H.~Georgi, {\it {Chiral Quarks and the Nonrelativistic Quark
  Model}},  {\em Nucl. Phys.} {\bf B234} (1984) 189--212.

\bibitem{Luty:1997fk}
M.~A. Luty, {\it {Naive dimensional analysis and supersymmetry}},  {\em Phys.
  Rev.} {\bf D57} (1998) 1531--1538,
  [\href{http://arxiv.org/abs/hep-ph/9706235}{{\tt hep-ph/9706235}}].

\bibitem{Cohen:1997rt}
A.~G. Cohen, D.~B. Kaplan, and A.~E. Nelson, {\it {Counting 4 pis in strongly
  coupled supersymmetry}},  {\em Phys. Lett.} {\bf B412} (1997) 301--308,
  [\href{http://arxiv.org/abs/hep-ph/9706275}{{\tt hep-ph/9706275}}].

\bibitem{Giudice:2007fh}
G.~F. Giudice, C.~Grojean, A.~Pomarol, and R.~Rattazzi, {\it {The
  Strongly-Interacting Light Higgs}},  {\em JHEP} {\bf 06} (2007) 045,
  [\href{http://arxiv.org/abs/hep-ph/0703164}{{\tt hep-ph/0703164}}].

\bibitem{Contino:2016jqw}
R.~Contino, A.~Falkowski, F.~Goertz, C.~Grojean, and F.~Riva, {\it {On the
  Validity of the Effective Field Theory Approach to SM Precision Tests}},
  {\em JHEP} {\bf 07} (2016) 144, [\href{http://arxiv.org/abs/1604.06444}{{\tt
  arXiv:1604.06444}}].

\end{thebibliography}\endgroup

\end{document}